\documentclass[journal=jctcce,manuscript=article]{achemso}

\usepackage[english]{babel}
\addto\captionsenglish{

}

\usepackage[sc]{mathpazo}        
\usepackage{units}               
\usepackage{xspace}
\usepackage{booktabs}            
\usepackage{dcolumn}             
\usepackage[table]{xcolor}
\usepackage{amsmath}
\usepackage{multirow}
\usepackage{numprint}
\usepackage{caption}
\usepackage{spreadtab}           
\STsetdecimalsep{.}              
\npdecimalsign{.}                
\usepackage[numbers,super,sort&compress]{natbib}
\usepackage{graphicx}            

\usepackage[official]{eurosym}   
\usepackage{url}
\definecolor{LinkCol}{cmyk}{1.00 0.00 0.57 0.30} 

\usepackage[%
colorlinks=true,
linkcolor=LinkCol,
urlcolor=LinkCol,
citecolor=LinkCol]{hyperref}

\newcommand{\gromacs}{GROMACS\xspace}
\newcommand{\nsday}{\nicefrac{ns}{d}\xspace}
\newcommand{\stern}{\mbox{$^\star$}}
\newcommand{\daggr}{\mbox{$^\dagger$}}
\newcommand{\ddagg}{\mbox{$^\ddagger$}}
\newcommand{\mal}{\mbox{$\times$}\xspace}

\newcommand{\cuEight}{\daggr}
\newcommand{\cuNine}{\stern}
\newcommand{\mpcdf}{\ddagg}

\newcommand{\Euro}{\euro\xspace}
\newcommand{\two}{\mbox{$\times$2}}
\newcommand{\three}{\mbox{$\times$3}}
\newcommand{\four}{\mbox{$\times$4}}

\newcommand{\mult}{\multicolumn{1}{l}}

\newcolumntype{L}{>{$}l<{$}}
\newcolumntype{C}{>{$}c<{$}}
\newcolumntype{R}{>{$}r<{$}}
\newcolumntype{S}{D{.}{.}{-1}}

\STautoround{2}

\FPset\fMEM{122}
\FPset\fRIB{1530}
\FPset\fVES{1.39}
\FPset\fBIG{55.5}

\FPset\eurBbBroadwell{1070}        
\FPset\eurBbSkylake{1250}          
\FPset\eurBoardBroadwell{225}      
\FPset\eurBoardSkylake{340}        
\FPset\eurBoardCoreiThree{230}     
\FPset\eurBoardEThree{175}
\FPset\eurChassis{210}

\FPset\eurCpuTenCoreBroadwell{535} 
\FPset\eurCpuFourCoreSkylake{285}  
\FPset\eurCpuTenCoreSkylake{580}   
\FPset\eurCpuEightCoreSkylake{400} 
\FPset\eurCpuFourCoreiSeven{250}   
\FPset\eurCpuFourCoreKaby{250}     
\FPset\eurCpuTwCoreSkylake{2800}   

\FPset\eurGtxTenEightyTi{610}      
\FPset\eurGtxTenEighty{420} 
\FPset\eurGtxTenSeventyTi{375}
\FPset\eurGtxTenSeventy{310} 
\FPset\eurRtxTwSeventy{450}
\FPset\eurRtxTwEighty{640}
\FPset\eurRtxTwEightyTi{1050}
\FPset\eurTeslaVIOO{8000} 

\FPset\eurRam{300}
\FPset\eurSsd{50}

\FPset\fOld{1.025}             

\newcommand{\MPIbpc}{Theoretical and Computational Biophysics, Max Planck
Institute for Biophysical Chemistry, Am Fassberg 11, 37077 G\"ottingen, Germany}
\newcommand{\KTH}{Center for High Performance Computing, KTH Royal Institute of Technology,
10044 Stockholm, Sweden}

\author{Carsten Kutzner}
\affiliation{\MPIbpc}
\email{ckutzne@gwdg.de}

\author{Szil\'ard P\'all}
\affiliation{\KTH}

\author{Martin Fechner}
\affiliation{\MPIbpc}

\author{Ansgar Esztermann}
\affiliation{\MPIbpc}

\author{Bert L. de Groot}
\affiliation{\MPIbpc}

\author{Helmut Grubm\"uller}
\affiliation{\MPIbpc}

\newcommand{\ManuscriptTitle}{More Bang for Your Buck:\\
Improved use of GPU Nodes for \gromacs 2018}

\title{\ManuscriptTitle}

\begin{document}

\hyphenation{off-loaded off-load off-loading Ge-Force over-clocked}

\begin{abstract}
We identify hardware that is optimal
to produce molecular dynamics trajectories on Linux compute clusters
with the \gromacs 2018 simulation package.
Therefore, we benchmark the \gromacs performance on a diverse set of compute nodes
and relate it to the costs of the nodes, which may include their lifetime costs for energy and cooling.
In agreement with our earlier investigation using \gromacs 4.6 on hardware of 2014,
the performance to price ratio of consumer GPU nodes is
considerably higher than that of CPU nodes.
However, with \gromacs 2018,
the optimal CPU to GPU processing power balance has shifted even more towards the GPU.
Hence, nodes optimized for \gromacs 2018 and later versions enable a significantly higher 
performance to price ratio than nodes optimized for older \gromacs versions.
Moreover, the shift towards GPU processing allows to
cheaply upgrade old nodes with recent GPUs,
yielding essentially the same performance as comparable brand-new hardware.
\end{abstract}


\section{Introduction}
Molecular dynamics (MD) simulation is a well established computational tool to investigate and
understand biomolecular function in atomic detail from a physics perspective.
A simulation system of a solvated molecule can comprise thousands 
to millions of atoms, depending on whether it is a small protein or
a large complex like a ribosome\cite{escidoc:1854600} or a viral shell.\cite{escidoc:1477658}
To derive the time evolution of atomic movements on biologically relevant
time scales, millions of time steps need to be computed.
For statistically significant results, this process is usually repeated many times
with varying starting conditions. 
Consequently, the investigation of a single biomolecular system can
easily occupy a number of modern compute nodes for weeks, whereas all simulation projects
of a typical research group performing MD calculations
will keep a medium-sized compute cluster running non-stop.

Whether the necessary cluster hardware is purchased by the department that uses it or
the services of a high performance computing (HPC) center are used, eventually
someone has to decide on what to buy.
This decision is not straightforward as the available hardware is quite diverse.
What node specifications should be used? Should they rather have many weak compute cores
or fewer strong ones? Are multi-socket nodes better than single-socket nodes? 
How many CPU cores are needed per GPU and what GPU type is optimal?
What about memory and interconnect?

All-rounder cluster nodes designed for many diverse software applications usually contain 
top-end CPUs, GPUs with high double precision floating point performance,
lots of memory, and an expensive interconnect. 
The result of meeting all these needs at once is 
a very low ratio of computation performance to node price for each individual application.
Our approach is completely opposite: maximized cost-efficiency by specialization.
We focus on a particular application, namely MD, and look for hardware that yields
the highest simulation throughput for a fixed budget, measured in total length of
produced trajectory over its lifetime.

The set of available MD codes for biomolecular simulations is diverse and includes, among others,
ACEMD,\cite{Harvey2009} 
Amber,\cite{salomon2013routine}
CHARMM,\cite{CHARMM2009}
Desmond,\cite{Bowers2006}
LAMMPS,\cite{Brown2012449}
NAMD,\cite{phillips2005scalable}
OpenMM,\cite{eastman2012openmm}
and \gromacs.\cite{Abraham2015}
We use \gromacs, because it is one of the fastest MD engines available,
widely-used, and freely available.

Our basic question is:
Given a fixed budget,
how can we produce as much MD trajectory as possible? Accordingly,
we measure simulation performances for representative biomolecular MD systems
and determine the corresponding total hardware price.
We do not aim at a comprehensive evaluation of currently available hardware,
we merely aim at uncovering hardware that has an exceptional performance to price (P/P) ratio,
which is the efficiency metric used in this study, for version 2018 of the \gromacs MD code.

As our study prioritizes the efficiency and total throughput of generating trajectories
(assuming plenty of concurrent simulations),
we do not consider use-cases where generating an individual trajectory as fast as possible is preferred.
Whereas the latter can be important, e.g.\ in exploratory studies,
faster simulations require strong scaling, which always comes at a cost due to the inherent
trade-off between simulation rate and parallel efficiency.
At the same time, running a large number of independent or weakly coupled simulations
is a widely used and increasingly important use-case.
Thanks to continuous advances in simulation performance,
many well-optimized MD engines like that of \gromacs 
have useful trajectory generation rates even without large-scale parallelization.
Additionally, problems that previously required the generation of a few long trajectories are nowadays
often addressed by ensemble methods that instead need a large number of shorter 
trajectories.\cite{Lundborg2018,chodera2014markov,buch2011,plattner2017,Chen438994}
Hence, large compute resources with a single fast interconnect are often not a must.
Instead, the ideal hardware in these instances
is one with fast ``islands'' and a more modest interconnect between these.
In fact, on such machines, results can in some cases be obtained as fast or faster,
and crucially more cost-effectively.
In our case, considering current hardware limitations and software characteristics,
for ultimate efficiency reasons, the fast ``islands'' are represented by a set of CPU cores and a single GPU.

In addition to the P/P ratio, when evaluating systems,
we take into account the two following criteria:
(i) energy consumption, as it is one of the largest contributors to trajectory production costs and
(ii) rack space, which is limited in any server room.
We cover energy consumption in a separate section of the paper, whereas
space requirements are implicitly taken into account by our hardware preselection;
We require an average packing density of at least one GPU per height unit U of rack space
for server nodes to be considered: a 4 U server with 4 GPUs meets the criterion.

In an earlier investigation using \gromacs 4.6 with 2014 hardware,\cite{escidoc:2180273}
we showed that the simulation throughput of an optimized cluster is
typically two to three times larger than that of a conventional cluster.
Since 2014, hardware has been evolving continuously and fundamental algorithmic enhancements 
have been made. Therefore, using the two exact same MD test systems,
we provide this update to our original investigation
and point the reader to the current hardware yielding the \emph{best bang for your buck} with 
\gromacs 2018.\cite{Abraham2015,Pall:2015}

We focus on hardware evaluation 
and not on how \gromacs performance can be optimized,
as that has already been extensively discussed.\citep{escidoc:2180273}
Most of the performance advise given in our original paper are still valid
if not stated otherwise.
Particular remarks for a specific \gromacs version are available in the online user guide at
\url{http://manual.gromacs.org/} in the section called \emph{Getting good performance from mdrun}.

\subsection{\gromacs load distribution schemes}
\gromacs uses various mechanisms to parallelize compute work over available resources
so that it can optimally benefit from the hardware's capabilities.\cite{Hess:2008tf, Abraham2015}
Processes sharing the computation of an MD system (these processes are called ranks) communicate via the 
Message Passing Interface (MPI) library,
while each rank can consist of several OpenMP threads.
Each rank can optionally have its short-range part of the Coulomb and van der Waals interactions
(the \emph{pair interactions})
calculated on a GPU, if present; this process is called offloading and is illustrated in Fig.~\ref{fig:offloading}.
The long-range part of Coulomb interactions is computed with the Particle Mesh Ewald (PME) method,\cite{Essmann:1995vj}
which can be performed on a subset of the available ranks for improved parallel efficiency.
Alternatively, from version 2018 on, PME can also be offloaded to a GPU.
On the lowest level of the parallelization hierarchy, SIMD (single instruction multiple data) 
parallelism is exploited in almost all performance-sensitive code parts.

\subsection{Summary of the 2014 investigation}
To illustrate the advancements, both on the implementation side as well as
on the hardware side, over the past five years,
we summarize the main points of our original investigation in the following.

In our original investigation,\cite{escidoc:2180273} we determined hardware prices
and \gromacs 4.6 performances for over 50 different node configurations 
built from 12 CPU and 13 GPU models.
In particular, we compared consumer GPUs with professional GPUs.
Professional GPUs like NVIDIA's Tesla and Quadro models 
are typically used for computer-aided design, computer-generated imagery, and in HPC.
Consumer GPUs, like the GeForce series, are mainly used for gaming. They are
much cheaper (up to about 1000~\euro\ net compared to several thousand Euro for the professional cards) and lack some of the
features the professional models offer, e.g.\ regarding memory and double precision floating-point performance.

\begin{table}[tbp]
\begin{center}
\caption{Specifications of the MD benchmarks.
Our principal benchmarks are done with atomistic MEM and RIB systems,
whereas for comparison, we also benchmark two coarse grain systems 
(VES and BIG, see section \ref{sec:martini}) using the Martini force field.}
\label{tab:systems}
\small
\begin{tabular}{lccccc} \toprule
MD system                    & MEM\cite{escidoc:599912}& RIB\cite{escidoc:1854600} & VES\cite{vesicleWeb}        & BIG                         \\ \midrule
\# particles                 & 81,743                  &         2,136,412         & 72,076                      &  2,094,812                  \\
system size (nm)             & 10.8 \mal 10.2 \mal 9.6 &         31.2$^3$          & 22.2 \mal 20.9 \mal 18.4    &  142.4 \mal 142.4 \mal 11.3 \\
time step (fs)               & 2                       &         4                 & 30                          &  20                         \\
cutoff radii (nm)            & 1.0                     &         1.0               & 1.1                         &  1.1                        \\
PME mesh spacing (nm)        & 0.12                    &         0.135             &  --                         &    --                       \\
\bottomrule
\end{tabular}
\end{center}
\end{table}
Our two main benchmark systems (that we continue to use in the present study) were 
an 80~k atom membrane protein embedded in a lipid bilayer surrounded by water and ions (MEM)
and a 2~M atom Ribosome system (RIB) solvated in water and ions, see Table~\ref{tab:systems}
for specification details.
On each hardware configuration, we determined the fastest settings for running a single simulation
of each system by performing a scan of the parameters controlling the parallelization settings,
i.e., the number of MPI ranks, the number of OpenMP threads, and
the number of separate PME ranks.

\begin{figure}
\begin{center}
\includegraphics[width=0.65\textwidth]{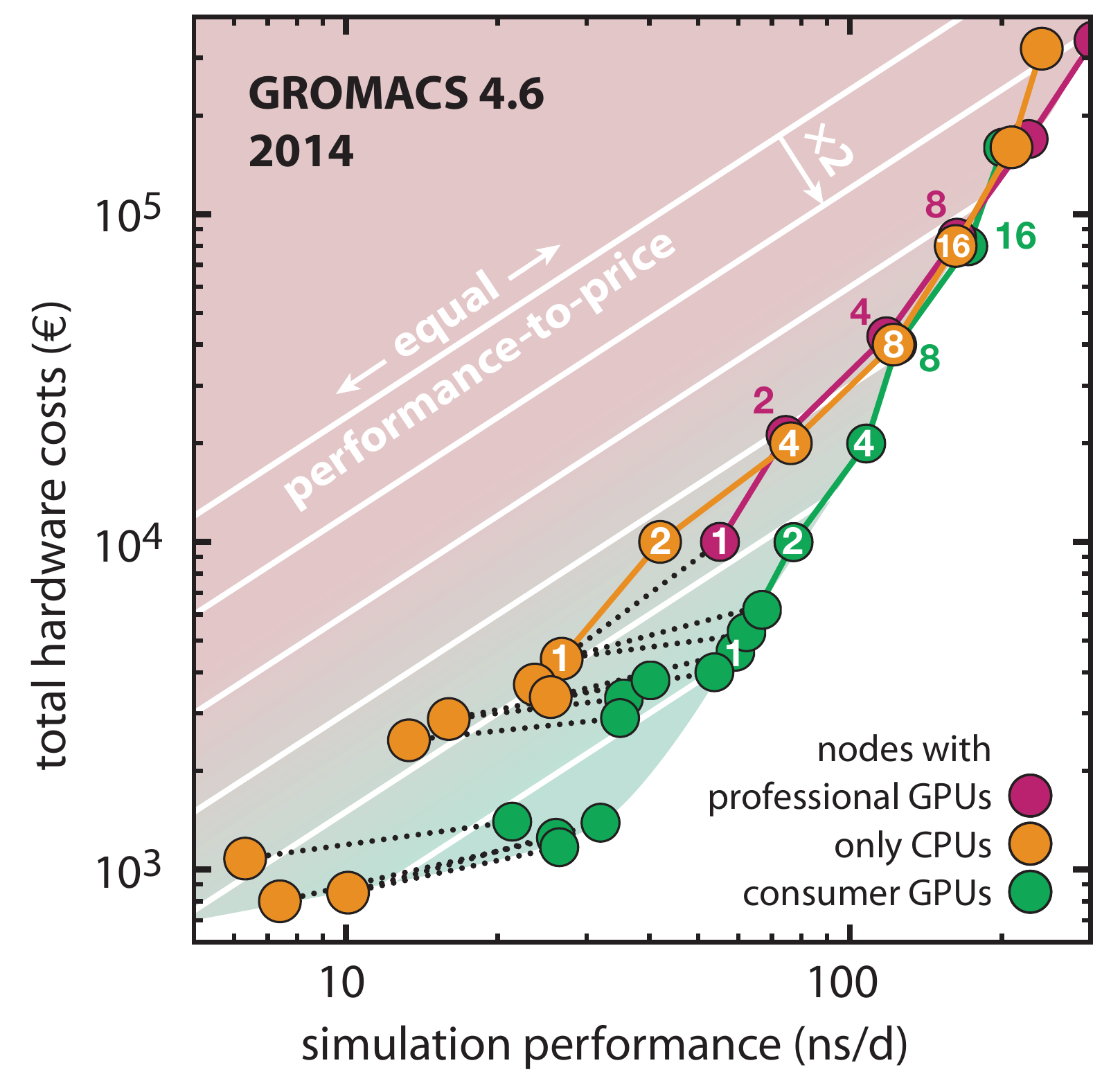}
\caption{Summary of our original investigation testing \gromacs 4.6 
on nodes built from hardware available in 2014.\cite{escidoc:2180273}
Hardware costs vs. MEM benchmark performance (circles) for three classes of node types:
CPU-only nodes (orange), nodes with professional Tesla GPUs (purple), and
nodes with consumer GeForce GPUs (green).
Dotted lines connect GPU nodes with their CPU counterparts.
Circles connected by colored lines denote a cluster built from that node type,
numbers in the circles denote how many of these nodes took part in the benchmark.
The white lines are isolines of equal P/P ratio.
Moving down from one isoline to the next increases the
P/P ratio by a factor of two 
(red shaded area = low P/P ratio, green shaded area = high P/P ratio). 
}
\label{fig:eval2014}
\end{center}
\end{figure}

We concluded from our investigation (Fig.~\ref{fig:eval2014}) that single CPU-only nodes
and nodes with professional GPUs have a comparably low P/P ratio,
whereas consumer GPUs improve the P/P ratio by a factor of 2--3.
Adding the first consumer GPU to a node yields the largest increase of the P/P ratio, 
whereas adding too much GPU power can also lower the P/P ratio
(for instance, 
compare the 2x E5-2680v2 nodes with 2 and 4 GTX980 GPUs in Fig.~6 of the original publication\cite{escidoc:2180273}).

Parallelizing a simulation over many nodes to increase the performance
leads to a dramatic decrease of the P/P ratio (Fig.~\ref{fig:eval2014}, top right corner).
For projects where the total amount of sampling is more important than the length of the
individual trajectories, it is therefore advisable to run many single-node simulations
instead of a few multi-node ones.

\subsection{Hardware and software developments 2014--2018}
\subsubsection*{Hardware developments}
\begin{figure}
\begin{center}
\includegraphics[scale=\commonScaleGpuPlots]{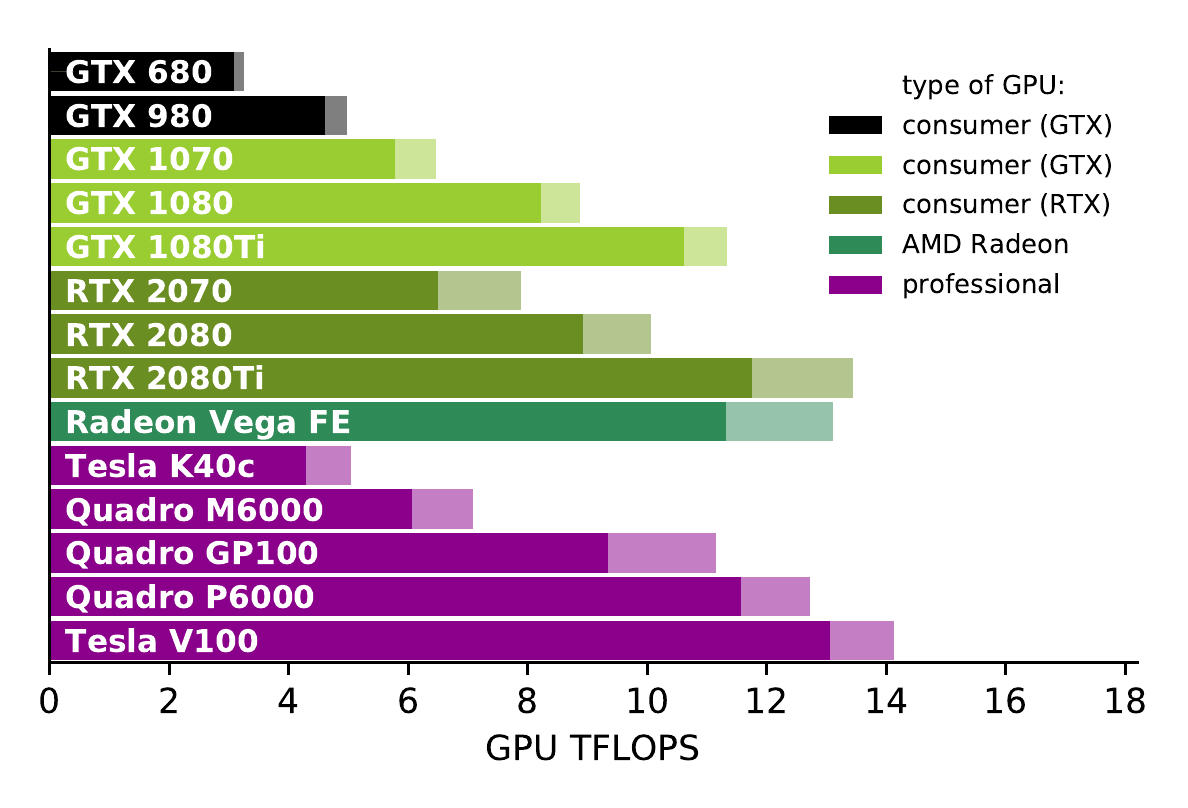}
\caption{Raw SP floating-point performance of selected GPU models
as computed from cores and clock rate 
(shaded area depicts FLOPS when running at boost/maximum application clock rate).
Consumer GPUs that were part of the 2014 investigation are depicted in black,
recent consumer GPUs in shades of green,
Quadro and Tesla professional GPUs in purple.
}
\label{fig:gpuFlops}
\end{center}
\end{figure}

\begin{figure}
\begin{center}
\includegraphics[scale=\commonScaleGpuPlots]{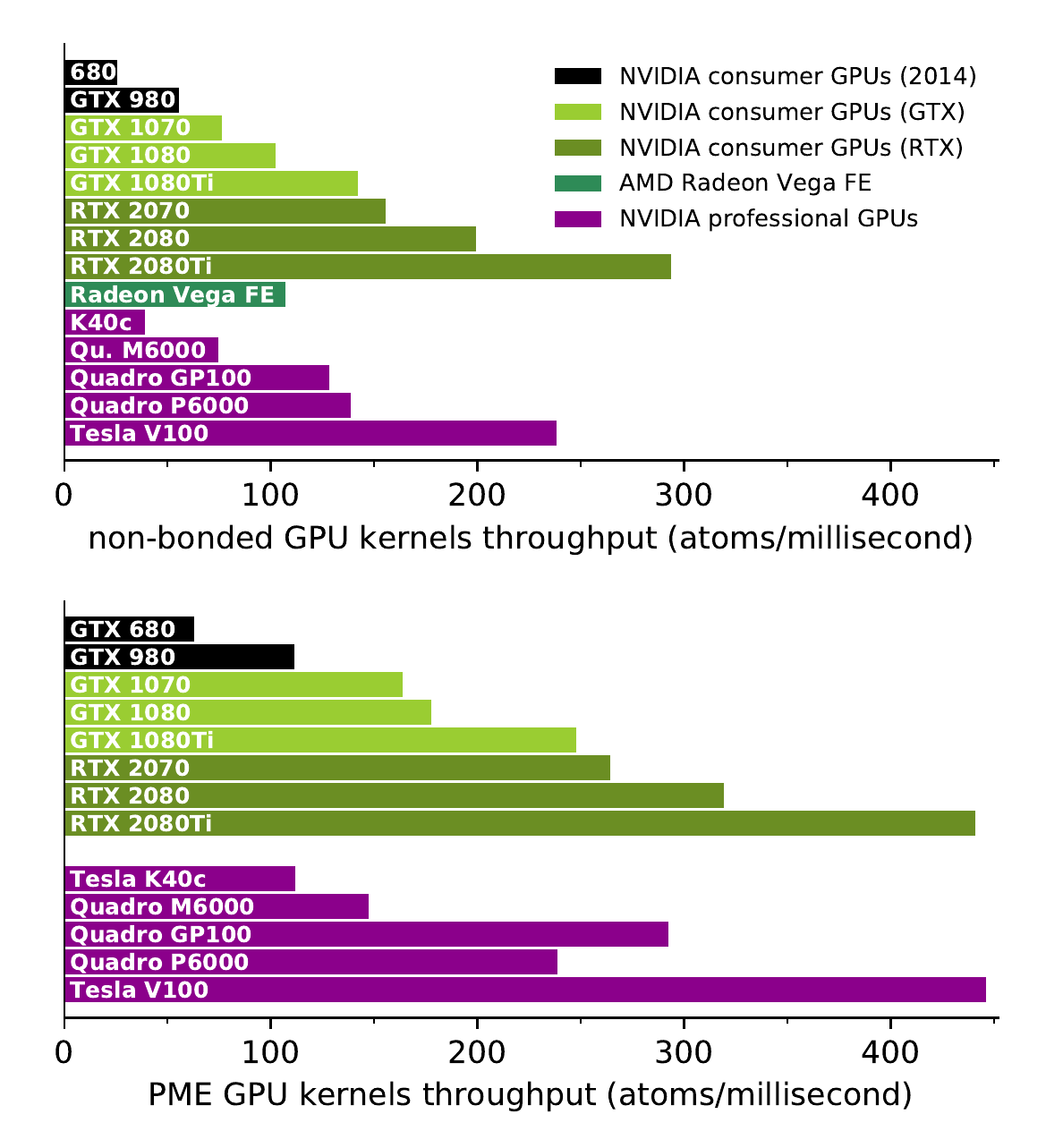}
\caption{
    Throughput of the GPU-offloaded computation:
    short-range non-bonded interactions (top panel) and PME long-range electrostatics (bottom panel)
    across GPU devices representing the various hardware generations and categories in this study.
    Coloring matches that of Fig.~\ref{fig:gpuFlops}.
    Throughput of computation is expressed as atoms per millisecond
    to aid comparing to the raw FLOPS.\\
    Measurements were done by profiling kernel execution (with concurrency disabled)
    of a run with a TIP3P water box of 384,000 atoms (cutoff radii $1.0$~nm, PME mesh spacing 0.125~nm, time step $2$~fs),
    chosen to allow comparing different GPUs with different scaling behavior
    each at peak throughput for both kernels.
    }
\label{fig:kernelPerf}
\end{center}
\end{figure}

Over the past five years, GPU compute power has significantly increased 
(compare Table~\ref{tab:NVIDIA} and Fig.~\ref{fig:gpuFlops}). 
The recent NVIDIA \emph{Turing} architecture GPUs (olive bars) are two to three generations ahead of
the \emph{Kepler} and \emph{Maxwell} architectures (black bars) we tested in 2014 and
have improved single precision (SP) floating point performance by more than threefold in this period.
This was enabled by a leap in semiconductor manufacturing technology,
shrinking transistors from the 28~nm process used in \emph{Kepler} and \emph{Maxwell}
to 12~nm on \emph{Volta} and \emph{Turing,} and increasing transistor count more than fivefold.
In contrast, during the same period,
CPU manufacturing has taken a more modest step forward from 22 to 14~nm.
However, effective MD application performance of GPUs has in some cases 
improved even more than what raw floating point performance would suggest, 
thanks to microarchitectural improvements making GPUs more efficient at and therefore better suited for general purpose compute.
As an example, consider the performance of the compute-bound non-bonded pair interaction kernel
(Fig.~\ref{fig:kernelPerf}, top panel). While the throughput increase between  
earlier GPU generations, like the Tesla K40 to Quadro M6000 to Quadro P6000, 
tracked the FLOP rate increase quite well (approximately $1.9$x for both),
the Tesla V100 shows $1.7$x improvement for only $1.1$x SP FLOP rate advantage
(see purple bar on Figs.~\ref{fig:gpuFlops} and \ref{fig:kernelPerf}).
Unlike the aforementioned professional GPUs with similar maximum power ratings, comparing consumer
GPUs is less straightforward. However, a similar pattern is still well illustrated when contrasting the
\emph{Pascal} generation GTX 1080Ti with the \emph{Turing} RTX 2080 GPU.
Despite the approximately 10\% lower FLOP rate as well as 10\% lower maximum power rating,
the 2080 is 40\% and 29\% faster in non-bonded and PME computation, respectively.
Across two generations of GPUs, we observe up to 6x and 4x performance improvement for the 
GPU-offloaded GROMACS compute kernels, the non-bonded pair interactions and PME electrostatics, respectively.
In contrast, while the theoretical FLOP rate of CPUs has increased by a similar rate as that of GPUs,
microarchitectural improvements like the wider AVX512 SIMD instruction sets translated into an only relatively modest
gain in application performance, even in thoroughly SIMD-optimized codes like \gromacs. 

This confluence of GPU manufacturing and architectural improvements has opened up a
significant performance gap between CPUs and GPUs, in particular for compute-bound applications
like MD, that do not heavily rely on double-precision floating point arithmetic.
Additionally, the performance per Watt advantage and the affordability of 
high performance consumer GPUs, thanks to the competitive computer gaming industry, have
further strengthened the role of GPUs in the MD community.
The application performance improvements on GPUs have also led to a shift in typical hardware balance,
important in applications that use offload-based heterogeneous parallelization,
which motivated developments toward further GPU offload capabilities of the \gromacs MD engine.

\begin{table}[tbp]
\begin{center}
\caption{Technical specifications of GPU models used in this study.\\
\footnotesize
Frequency and SP FLOP for professional NVIDIA GPUs are based on the default and maximum ``application clocks,''
while for the consumer NVIDIA and AMD GPUs are based on the published base and boost clocks.
Note that the NVIDIA GeForce GPUs will often operate at even higher clocks under compute workloads than those indicated by 
the boost clock (even without factory overclocking), especially in well-cooled environments.
For GPUs that were available in 2018, the last column lists the approximate net price in 2018.
Note that only the NVIDIA cards can execute CUDA code, whereas on AMD cards OpenCL can be used.
}
\label{tab:NVIDIA}
\small
\STautoround{1}

\begin{spreadtab}{{tabular}{ll c r c r n{2}{1} c n{2}{1} c}}
\toprule
@model      & @manu-    & @archi-  & @\text{compute}&\multicolumn{3}{c}{@base -- boost} &\multicolumn{3}{c}{@SP TFLOPS}                                                &@\text{$\approx$ price}        \\
@           & @facturer & @tecture & @\text{units}  &\multicolumn{3}{c}{@clock (MHz)}   &&&                                                                            &@\text{(\euro\ net)}           \\ \midrule
\multicolumn{10}{l}{@NVIDIA consumer GPUs:}\\
@GTX  680   & @NVIDIA   & @Kepler  &  1536          &      1006    &@-- &     1058      & \STcopy{v12}{[-4,0]*[-3,0]/500000} &@-- & \STcopy{v12}{[-6,0]*[-3,0]/500000} &@   --                         \\
@GTX  980   & @NVIDIA   & @Maxwell &  2048          &      1126    &@-- &     1216      &                                    &@-- &                                    &@   --                         \\
@GTX 1070   & @NVIDIA   & @Pascal  &  1920          &      1506    &@-- &     1683      &                                    &@-- &                                    &@  \eurGtxTenSeventy           \\ 
@GTX 1070Ti & @NVIDIA   & @Pascal  &  2432          &      1607    &@-- &     1683      &                                    &@-- &                                    &@  \eurGtxTenSeventyTi         \\ 
@GTX 1080   & @NVIDIA   & @Pascal  &  2560          &      1607    &@-- &     1733      &                                    &@-- &                                    &@  \eurGtxTenEighty            \\
@GTX 1080Ti & @NVIDIA   & @Pascal  &  3584          &      1480    &@-- &     1582      &                                    &@-- &                                    &@  \eurGtxTenEightyTi          \\ 
@RTX 2070   & @NVIDIA   & @Turing  &  2304          &      1410    &@-- &     1710      &                                    &@-- &                                    &@  \eurRtxTwSeventy            \\
@RTX 2080   & @NVIDIA   & @Turing  &  2944          &      1515    &@-- &     1710      &                                    &@-- &                                    &@  \eurRtxTwEighty             \\
@RTX 2080Ti & @NVIDIA   & @Turing  &  4352          &      1350    &@-- &     1545      &                                    &@-- &                                    &@  \eurRtxTwEightyTi           \\ \midrule
\multicolumn{10}{l}{@AMD GPUs:}\\
@Radeon Vega 64 & @AMD  & @Vega    &  4096          &      1247    &@-- &     1546      &                                    &@-- &                                    &@  390                         \\
@Radeon Vega FE & @AMD  & @Vega    &  4096          &      1382    &@-- &     1600      &                                    &@-- &                                    &@  850                         \\ \midrule 
\multicolumn{10}{l}{@NVIDIA professional GPUs:}\\
            & \SThidecol \SThiderow\\
@Tesla K40c  & @NVIDIA  & @Kepler  &  2880          &       745    &@-- &      875      &   4.3  &@-- & 5.0                                                            &@  --                          \\ 
@Tesla K80   & @NVIDIA  & @Kepler  &  4992          &       562    &@-- &      875      &   5.6  &@-- & 8.7                                                            &@  --                          \\
@Quadro M6000& @NVIDIA  & @Maxwell &  3072          &       988    &@-- &     1152      &   6.1  &@-- & 7.1                                                            &@  --                          \\
@Quadro GP100& @NVIDIA  & @Pascal  &  3584          &      1303    &@-- &     1556      &   9.3  &@-- & 11.1                                                           &@  --                          \\
@Quadro P6000& @NVIDIA  & @Pascal  &  3840          &      1506    &@-- &     1657      &   11.6 &@-- & 12.7                                                           &@  4600                        \\ 
    @Tesla V100 & @NVIDIA   & @Volta   &  5120      &      1275    &@-- &     1380      &   13.6 &@-- & 14.1                                                           &@  \eurTeslaVIOO               \\ 
    \bottomrule
\end{spreadtab}

\end{center}
\end{table}

\FPset\commonScaleGpuPlots{1.1}

\begin{figure}
\begin{center}
\includegraphics[scale=\commonScaleGpuPlots]{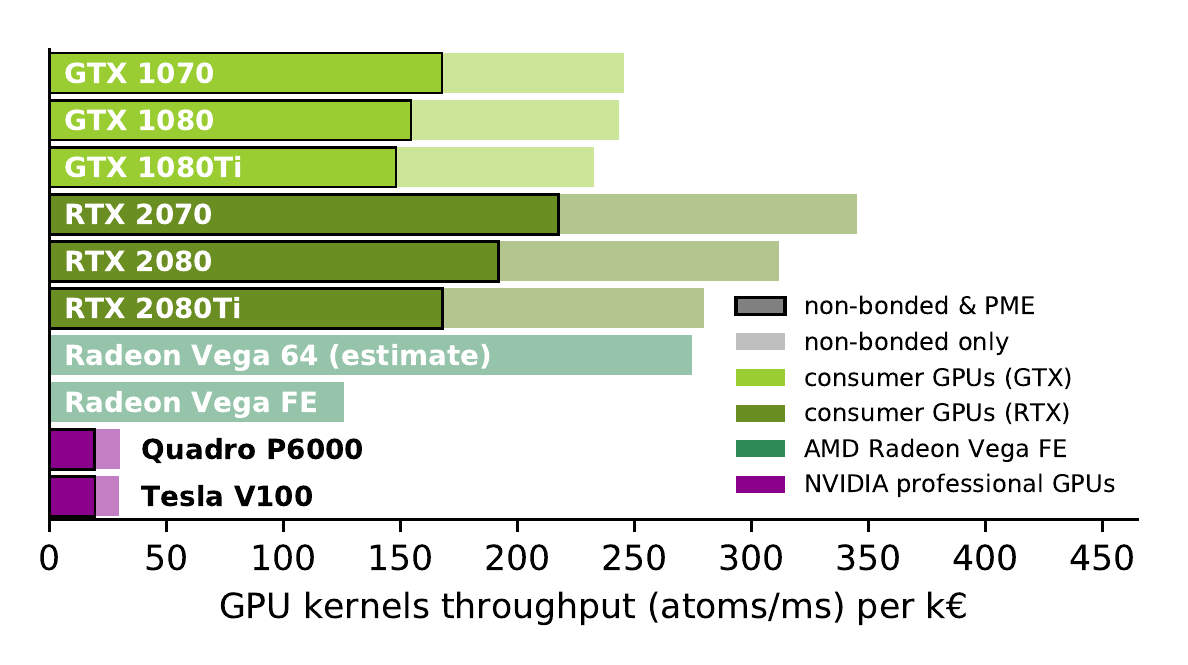}
\caption{Proxy metric for P/P ratio of selected GPU models,
computed as \gromacs 2018 GPU kernel performance divided by price
given in Table~\ref{tab:NVIDIA}. 
Light/shaded bars of each row show P/P derived for timings of non-bonded kernels only, 
dark bars show P/P derived from total timings of non-bonded and PME kernels.
The P/P of the AMD Vega 64 is estimated from the measured throughput of AMD Vega FE but the price of Vega 64.}
\label{fig:gpuPerf2Price}
\end{center}
\end{figure}

\subsubsection*{Software developments}
On the software side, there have been continuous improvements throughout the four major \gromacs releases
between our previous study and the present work: 5.0 (June 2014), 5.1 (August 2015), 2016 (August 2016), and 2018 (January 2018).
These releases yielded advances in algorithms (improved pair interaction buffer estimates),
SIMD parallelization (in PME, improved bonded, constraints, and new update kernels) and kernel optimizations (improved force accumulation),
as well as widespread multi-threading optimizations (like sparse summation for per-thread bonded force outputs).
Efforts in designing code to increase on-node parallelism
(wider SIMD units, higher core counts, and more accelerator-heavy compute nodes)
both aim at fueling performance improvements throughout the years making better use of existing hardware
but also at preparing the code for hardware evolution, an investment that promises future benefits.

Two significant improvements allowed the 2018 release to take further leaps in performance.
First, the \emph{dual pair list} extension of the cluster pair algorithm\cite{Pall2013} was developed with two goals:
reducing the computational cost of pair search (and domain decomposition) and facilitating
optimal simulation performance without manual parameter tuning.
The dual pair list algorithm enables retaining the pair list far longer while avoiding non-bonded computation overheads.
This is achieved by building a larger pair list less frequently (every 100--200 MD steps) using a
suitably longer buffered interaction cutoff. By using a frequent \emph{pruning} step based on a short buffered cutoff,
a smaller pair list with a lifetime of typically 5--15 MD steps (value auto-tuned at runtime) is obtained,
which is then used in evaluating pair interactions. 
Less frequent pair search significantly reduces the cost of this expensive computation
whereas the list pruning avoids introducing overheads of extra pair interaction evaluations due to a large buffer.
The added benefit is that tuning the search frequency to balance these two costs for optimal performance
is no longer needed.

\begin{figure}[tbp]
\begin{center}
\includegraphics[width=0.8\textwidth]{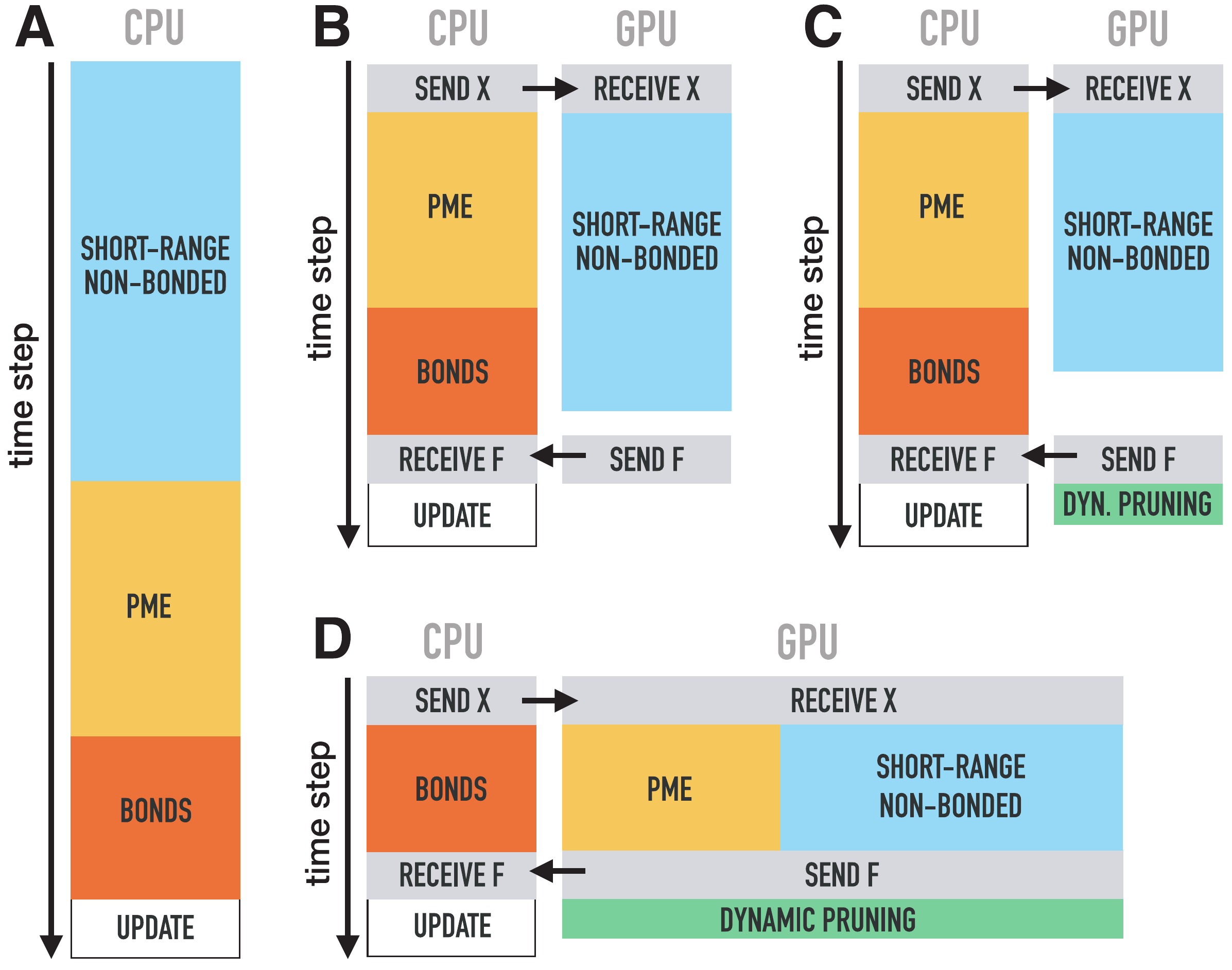}
\caption{
Comparison of different offloading schemes employed by
\gromacs 4.6, 5.x, and 2016 (B) and \gromacs 2018 (C, D).
Differently colored boxes represent the different main force computation parts of a typical MD step,
whereas grey boxes represent the CPU-GPU data transfers.
By offloading compute-intensive parts of the MD step
and using algorithms to optimize concurrent CPU-GPU execution,
the wall-time required by a time step (black vertical arrow) is decreased.\\
{\bfseries A:}
Without GPUs, the short-range non-bonded interactions (blue), PME (orange), 
and the bonded interactions (red) are computed on the CPU.
Once all forces have been derived, the atomic positions are updated (white).\\
{\bfseries B:}
Since version 4.6, GPU(s) can compute the non-bonded forces,
while the CPU(s) do PME and the bonded forces.
As a result, the wall clock time per MD step is significantly shortened,
at the small expense of having to communicate positions {\bfseries x} and forces {\bfseries F}
between CPU and GPU (grey).\\
{\bfseries C:}
Version 2018 introduced the dual pair list algorithm,
which a) reduces the number of short-range non-bonded interactions that are calculated, and b) reduces
the frequency of doing pair search on the CPU (not shown here).
There is no computational overhead added, as the dynamic list pruning (green box)
happens on the GPU(s) while the CPU updates the positions (white).\\
{\bfseries D:}
Since version 2018, also PME can be computed on a GPU,
further reducing the wall clock time per MD step,
if enough GPU processing power is available. 
}
\label{fig:offloading}
\end{center}
\end{figure}

The second major improvement is that, while \gromacs versions 4.6, 5.x, and 2016 could offload 
only the short-range part of Coulomb and van der Waals interactions to the GPU,
in the 2018 release
the PME mesh part can also be offloaded to CUDA enabled devices (Fig.~\ref{fig:offloading}).
By enabling offload of more work, the computational balance within \gromacs can be shifted
such that it exploits the shift in hardware balance in recent years. This improvement makes it possible to
efficiently utilize nodes with more and stronger GPUs
and it enables significantly higher P/P ratios than prior versions on recent hardware.
At the same time, sticking to the offload-based parallelization approach with improvements focused on
both CPU and GPU is still important and has two major benefits:
(i) It makes sure that nearly all of the versatile \gromacs features
remain supported with GPU acceleration (as opposed to limiting use of GPUs to only the subset of
common features ported to GPUs).
(ii) Additionally, offload allows minimizing the risk of vendor lock-in,
which is not negligible when the hardware of the dominant manufacturer can only be employed (with meaningful performance)
using proprietary tools and non standards-based tools and programming models.

As PME offloading with OpenCL will only be supported starting from the 2019 release,
we did not include AMD GPUs in our benchmark study yet.
However, we expect that recent AMD GPUs will be competitive against
similarly priced NVIDIA consumer GPUs, as suggested by both the non-bonded kernel performance
(see Fig.~\ref{fig:kernelPerf}) and by the proxy P/P ratio (Fig.~\ref{fig:gpuPerf2Price}).

The offloading approach
works best at a balanced GPU/CPU compute power ratio, 
i.e.\ if the ratio is adapted to the typical requirements of \gromacs simulations (Fig.~\ref{fig:offloading}).
With our benchmarks and our hardware choices, we aim to determine this optimum.
The more compute work is offloaded,
the more this balance is shifted towards the GPU,
which enables higher P/P ratios if GPUs are cheap.
The switch to \gromacs 2018 shifted the optimal CPU/GPU balance significantly towards the GPU,
as shown in the following. 

\subsubsection*{Assembling optimal GPU nodes for \gromacs 2018}
\begin{figure}
\begin{center}
\includegraphics[width=\textwidth]{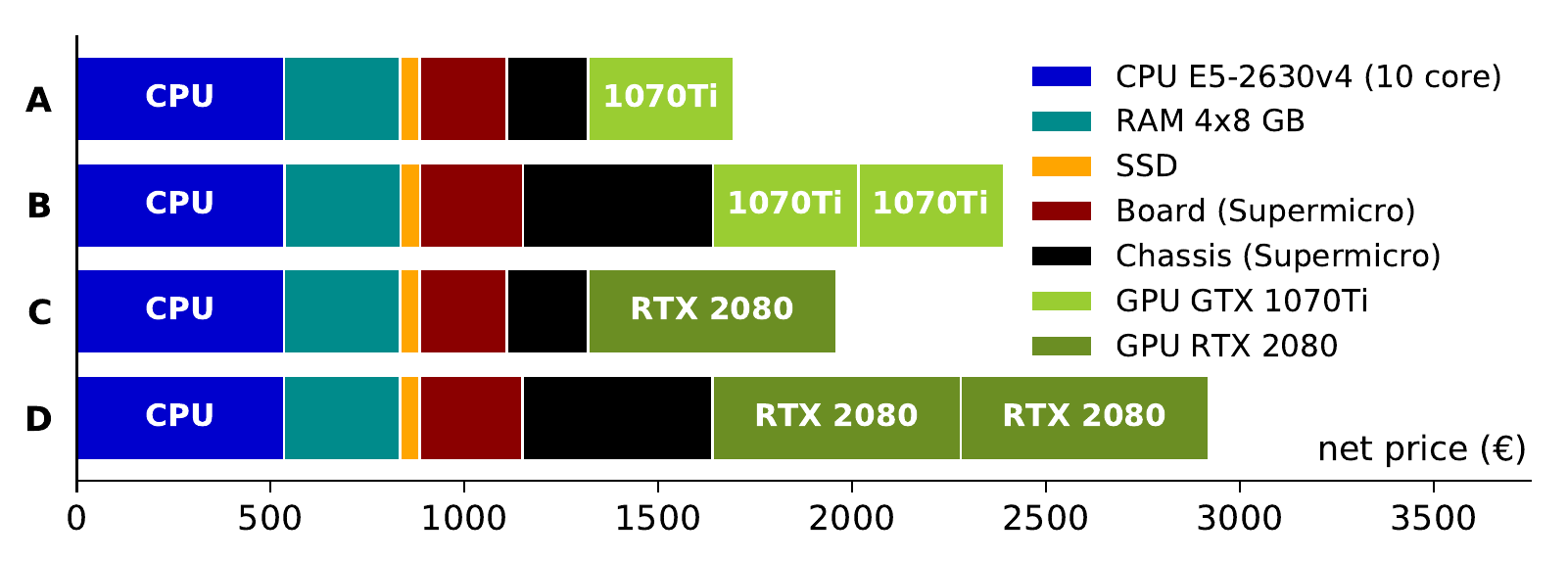}
\caption{Breakdown of node costs into individual components for four
exemplary node types.
Costs for CPU, RAM, and SSD disk are identical for nodes A--D, 
however, with two GPUs (B, D) a chassis with a stronger power supply has to be used,
making the node significantly more expensive.
}
\label{fig:nodeCosts}
\end{center}
\end{figure}

Choosing the right hardware components to build a node with a competitive P/P ratio
is a puzzle on its own (Fig.~\ref{fig:nodeCosts}).
Let us for a moment focus on the proxy P/P ratio of the GPUs only (Fig.~\ref{fig:gpuPerf2Price}).
Considering raw \gromacs GPU kernel throughput, of the \emph{Pascal} architecture GPUs,
the 1080 offers the highest P/P ratio,
whereas of the \emph{Turing} GPUs, the 2070 performs best.
However, considering GPUs with optimal P/P ratio only is not always be the best solution,
as a node is often more expensive the more GPUs it can accommodate
(see Fig.~\ref{fig:nodeCosts} for an example).
As a result, to optimize for the combined P/P ratio of a whole node, it typically is better to choose
a consumer GPU with a lower P/P ratio but higher performance.

Apart from the Gold6148F \mal 2 node with two Tesla V100's
(Table~\ref{tab:numbers2018} bottom, and Fig.~\ref{fig:costsVsPerf} top),
we did not build and benchmark any nodes with professional GPUs for two reasons:
(i) A Tesla GPU already costs more than most of the tested nodes including their consumer GPU(s).
(ii) For MD simulations with GROMACS,
the added benefit of using professional GPUs is marginal (ECC reliability, warranty)
and consumer models with comparable application performance generally exist.

\section{Methods}

For the main part of this study, the same two benchmark input files as in our 2014 investigation were used (Table~\ref{tab:systems}),
to facilitate the comparison between new and old hard- and software.

\subsection{Software environment}
Benchmarks done for evaluating \gromacs developments (Section \ref{sec:perf_developments})
used the latest release of each version.
All other benchmarks have been performed using \gromacs 2018, 
with \texttt{AVX2\_256} SIMD instructions switched on at compile time for
Intel CPUs and \texttt{AVX2\_128} SIMD instructions for AMD CPUs
(with the exception of Table~\ref{tab:upgrade}, where \texttt{AVX\_256} was
used reflecting the hardware capabilities).
Additionally, as version 4.6 did not have SIMD kernel support for the \texttt{AVX2\_256}
instruction set, here we used the \texttt{AVX\_256} build with adding the \verb+-mavx2 -mfma+
compiler optimization flags to allow \texttt{AVX2} instructions to be generated.

On nodes with no or a single GPU,
\gromacs' built-in thread-MPI library was used,
whereas on multi-GPU nodes Intel MPI 2017 was used.
OpenMP support was always enabled.
Additionally, \gromacs was linked against the portable hardware locality library \texttt{hwloc}\cite{broquedis2010hwloc} version 1.11.
On nodes without GPUs, the FFT needed by PME is calculated on the CPU.
Therefore, we employed FFTW 3.3.7,\cite{frigo2005design} 
compiled with GCC 4.8 using the options \verb+--enable-sse2 --enable-avx --enable-avx2+
that are recommended for best \gromacs performance.
On GPU nodes, CUDA cuFFT was automatically used with PME offloading.

All hardware was tested in the same software environment by booting the nodes from a common image
with Scientific Linux 7.4 as operating system 
(except the Gold6148/V100 nodes, which ran SLES 12p3, and the
\gromacs evaluation benchmarks, which ran on nodes with Ubuntu server 16.04.)

\gromacs was compiled either with GCC 5.4 and CUDA 8.0 or with GCC 6.4 and CUDA 9.1 for the main study,
while in the \gromacs evaluation section GCC 7.3 and CUDA 10 was used.

\subsection{Impact of compiler choice and CUDA version}
To assess the impact of the chosen GCC/CUDA combination on the measured performances,
we ran MEM benchmarks on identical hardware, but with both CUDA/GCC combinations.
On a node with two E5-2670v2 CPUs plus two GTX 1080Ti GPUs, our MEM benchmark
runs consistently faster with GCC6.4/CUDA9.1 over GCC 5.4/CUDA8,
by 3.5\% (average over 10 runs).
On an E3-1270v2 CPU with GTX 1070, the factor is about 4\%, whereas
on a workstation with E5-1650v4 and GTX 980 it is about 1.5\%.
To correct for this effect when comparing hardware,
the performances measured with the older GCC/CUDA combination have been
multiplied with the factor \fOld.

The performance difference between using CUDA 9.1 and CUDA 10.0 was determined
in a similar manner and as it turned out to be less than 0.5\%, we did not correct for this small effect.

\subsection{Benchmarking \gromacs performance evolution}
For \gromacs evaluation benchmarking (Figs.~\ref{fig:perfEvolution} and \ref{fig:perfVsCores}),
data collection protocols tailored for characterizing performance
of the various code versions were used.
All runs were carried out using a GPU attached to the PCI bus of the CPU employed
(or the first CPU where the master thread of the run was located when both CPUs were used in Fig.~\ref{fig:perfVsCores}).
Two CPU threads per core were used, profiting from HyperThreading with threads pinned.

For the evaluation of performance as a function of CPU cores per GPU (Fig.~\ref{fig:perfVsCores}),
we would ideally use CPUs that only differ in the number of cores and are identical otherwise.
We mimicked such a scenario with a single CPU model by using just a part of its available cores.
\renewcommand{\thefootnote}{*}
However, as modern CPUs, when only partially utilized, can increase the clock frequency of the busy cores,
the comparison would be unfair. We therefore made sure that the cores not used by our benchmark
were kept busy with
a concurrent CPU-only \gromacs run (using the same input
system as the benchmark), so that
approximately 
the same clock frequency is used independent of how many cores the benchmark runs on.\footnote{
    Note that with this protocol the last level cache is shared by the two runs co-located
    on the CPU. Hence, measurements are not equivalent with turning off CPU cores 
    not intended to be used
    and fixing a constant CPU frequency across all active cores, an alternative which
    would provide benchmarks for a small number of cores with an unrealistic amount of cache.}
\gromacs evaluation benchmarks  were repeated three times, with averages shown.

Lastly, all compile and runtime settings, other than the ones tested, were left at their default
or auto-tuned values (including pair search frequency and CPU-GPU load balancing).

\subsection{Measuring the performance of a node}
The performance of different node types can be assessed in various ways.
Moreover, different benchmark procedures lead to different hardware rankings.
Our requirements on the benchmarking procedure and the resulting hardware ranking were: 
(i) The benchmarks should closely mimic the intended hardware use.
(ii) Aggregation of compute power (e.g.\ combining hardware components
of two nodes into a single node), which may offer price and rack space savings,
should not be penalized. 

Our motivation for the second requirement is the following:
Assume you compare a) two single-socket nodes with CPU X and GPU Y each, 
to b) a dual-socket node with two CPUs of type X and two GPUs of type Y. 
The aggregate performance of a) and b) is expected to be identical, 
as two independent simulations can always run on b).

A benchmark protocol matching both requirements is:
running $N$ simulations on nodes with $N$ GPUs
in parallel, each using $1/N$ of the available CPU cores or hardware threads,
and reporting the aggregate performance, i.e.\ the sum of the performances of the individual simulations.
This protocol can easily be executed by using the \gromacs \texttt{-multidir} command line argument.
As in the initial phase, the load balancing mechanisms have not yet found their optimum,
we excluded the first $n$ time steps from the performance measurements
that were run for a total of $n_\text{tot}$ time steps.
For the MEM and VES benchmarks, we used $n = 15,000$ and $n_\text{tot} = 20,000$,
whereas for the RIB and BIG benchmarks, we used $n = 8,000$ and $n_\text{tot}=10,000$.

On single-socket nodes with one GPU, 
using a single rank with as many OpenMP threads as available cores (or hardware threads) is usually fastest,
as one avoids the overhead introduced by MPI and domain decomposition.\cite{Pall:2015,escidoc:2180273} 
Therefore, the single-GPU benchmarks of the old investigation and the present study are comparable, 
whereas on multi-GPU nodes, the new benchmark protocol is expected to
yield a higher aggregate performance than the single-simulation performance that was
measured in 2014.

\subsection{Power consumption measurements}
We measured the power consumption of a node with the Linux programs
\texttt{ipmi-sensors} version 1.5.7 (\url{http://www.gnu.org/software/freeipmi/})
and
\texttt{ipmitool} version 1.8.17 (\url{https://github.com/ipmitool/ipmitool}).
On some nodes, e.g.\ on the Ryzen workstations that do not support reading out the power
draw via \texttt{ipmi}, we used a VOLTCRAFT Power Monitor Pro multimeter.
In all cases, we computed the average of 60 separate power readings with one second time difference each.
During the power measurements, the RIB benchmark was running with the same settings
that were used to derive the performance. 

\subsection{Coarse grain simulations with Martini}
\label{sec:martini}
In order to check to what extent the results from the atomistic benchmarks are transferable to coarse grain
simulations, we added benchmarks that use the
Martini\cite{marrink2004coarse, marrink2007martini} force field. 
With Martini, the particle density is lower than for all-atom force fields and the electrostatic
interactions are usually not calculated via PME. Therefore, it is unclear whether the
hardware optimum for coarse grain simulations is the same as for atomistic simulations.

To facilitate the comparison with the atomistic systems, we chose coarse grain systems with
a similar number of particles, respectively, see Tab.~\ref{tab:systems}.
The small Martini benchmark system ``VES'' is a POPE vesicle in water 
and with a total of 72,076 particles, comparable in size with the MEM atomistic benchmark.
It is one of the example applications available for download 
at \url{www.cgmartini.nl}.\cite{vesicleWeb}
The large Martini benchmark system ``BIG'' was created from the mammalian plasma membrane
example system.\cite{membraneWeb}
To arrive at the final large benchmark system, a patch of 2 x 2 copies in x- and y- directions
of the original membrane was created with \texttt{gmx genconf}, resulting
in a final membrane with 2,094,812 particles, thus comparable in size to the
RIB atomistic benchmark.

Following the suggestions for Martini simulations with GPUs, 
we used the \emph{New-RF} set of simulation parameters.\cite{de2016martini,benedetti2018comment}
As with version 2018 the dual pair list algorithm was introduced (see Fig.~\ref{fig:offloading}C),
we increased the neighbor searching interval from 20 (as used in the \emph{New-RF} parameter set) 
to 50 steps (with an inner pair list lifetime of 4 steps) for improved performance.

\section{Results}

\subsection{Evaluation of GROMACS performance developments}
\label{sec:perf_developments}

\begin{figure}[tb]
\begin{center}
\includegraphics[width=0.8\textwidth]{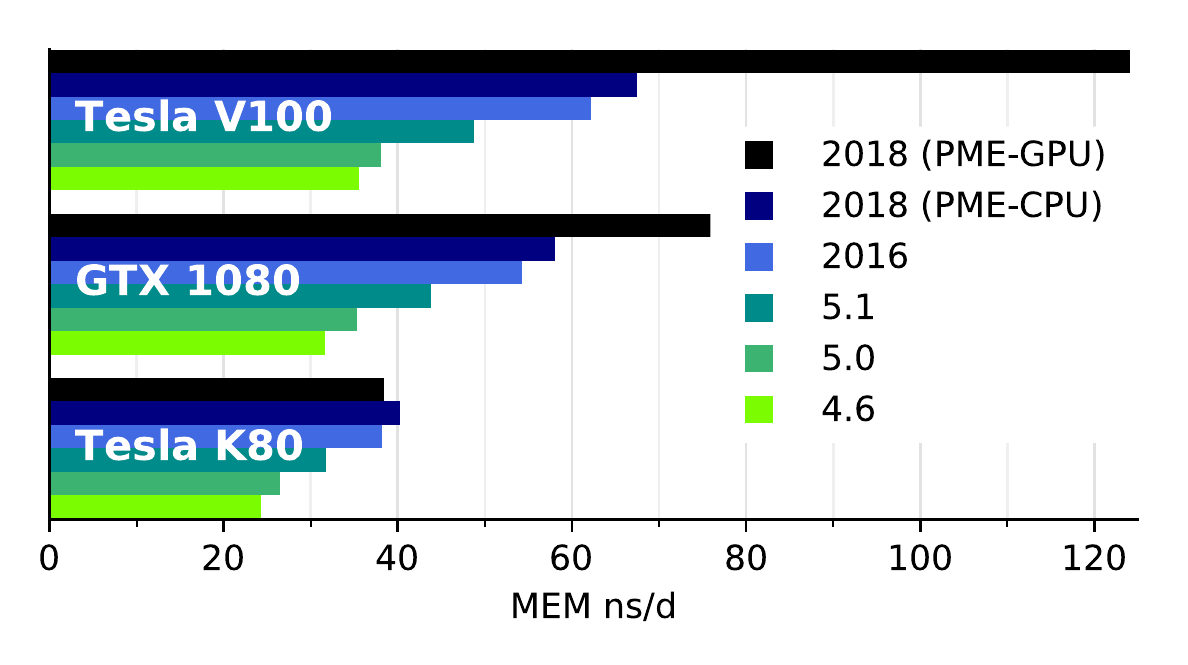}
\caption{Evolution of the \gromacs performance on GPU nodes for versions 4.6 -- 2018.
The short-range non-bonded interactions were offloaded to the GPU in all cases,
whereas for version 2018, also the PME mesh contribution can be offloaded (topmost black bars). 
Tesla V100 and GTX 1080 GPUs were mounted in a node with two Xeon E5-2620v4 processors (2x 8 cores).
The Tesla K80 GPU was mounted in a node with two Xeon E5-2620v3 processors (2x 6 cores).
}
\label{fig:perfEvolution}
\end{center}
\end{figure}

As shown in Fig.~\ref{fig:perfEvolution},
thanks to the algorithmic improvements and optimizations described earlier,
across the initial four releases since our previous study, between versions 4.6 and 2016,
simulation performance improved by up to 65\% on previous-generation hardware (e.g.~Tesla K80)
and by as much as 75--90\% on more recent hardware (e.g.~GTX1080 and Tesla V100).
Between the 2016 and 2018 versions with PME on the CPU, we measured a 6\%--9\% performance increase
(light and dark blue bars in Fig.~\ref{fig:perfEvolution}), which is largely due to the dynamic pruning algorithm. 
We expect this advantage to grow even larger with future GPU hardware because
the faster the GPU-offloaded computation gets, the larger the benefit of this algorithm will be.
For the given benchmark systems and server setups, we observe additional PME offload improvements
of 35--84\% when using recent GPUs (see black bars in Fig.~\ref{fig:perfEvolution}).
At the same time, on the legacy hardware setup, offloading to the older generation
Tesla K80 leads to slowdown. To better understand this performance change,
we explore the performance characteristics of the heterogeneous
PME-offload code in \gromacs 2018 in the following.

\begin{figure}
\begin{center}
\includegraphics[width=\textwidth]{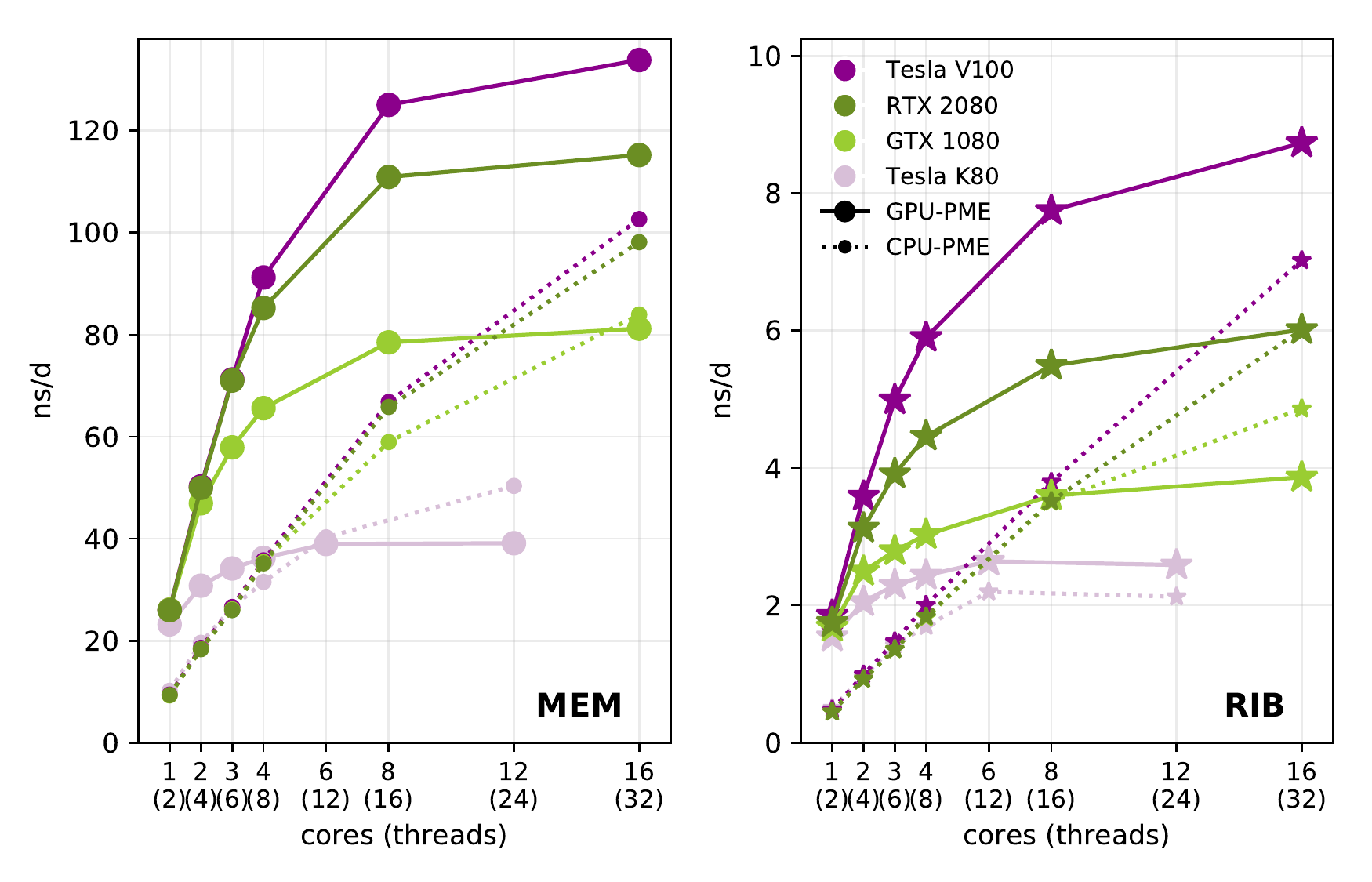}
\caption{\gromacs 2018 performance as a function of CPU cores used per GPU.
The GTX 1080, RTX 2080, and Tesla V100 cards were installed in server nodes along 
Xeon E5-2620v4 processors (2x 8 cores), whereas the Tesla K80 was installed in a node with two E5-2620v3 processors (2x 6 cores).
Solid lines illustrate performance with PME offloaded to the GPU, whereas
dotted lines with smaller symbols indicate performance with PME computed on the CPU (cores indicated on the horizontal axis).
}
\label{fig:perfVsCores}
\end{center}
\end{figure}

Our evaluation benchmarks were carried out on servers representing GPU-dense setups,
consisting of Xeon CPUs with rather modest performance combined with fast accelerators.
These are traditionally challenging for GROMACS as illustrated by the strong dependence
of the performance on the number of cores used (indicated by the dotted lines in Fig.~\ref{fig:perfVsCores}).
One of the main performance goals of the GROMACS 2018 development was to reach a close to the
peak simulation rate of the previous offload scheme (non-bonded only) on balanced hardware,
but with only a few CPU cores accompanying the GPU.
Of our benchmarked hardware setups, the GTX 1080 (light green curves in Fig.~\ref{fig:perfVsCores})
combined with about 12--14 cores of the E5-2620v4 processors could be considered a balanced setup.
For the two systems, with only four cores per GPU, the PME-offload feature allows reaching 80\% and 90\%
of the peak with no PME offload, respectively.

Whereas the 2016 release required many fast cores to achieve a good load balance between CPU and GPU,
with the 2018 release, in most cases, only 4--6 slower (typical server) cores as the ones in our benchmarked systems
are sufficient to reach $>$80\% of the peak simulation performance (e.g.~as obtained with all 16 cores of a node here).
Workstations typically have a few cores, but these are fast.
In contrast, servers often have more but slower cores than workstations.
To compare the raw CPU processing power of both node types, we consider
the ``core-GHz'', i.e.~the number of cores multiplied with the clock frequency.
We determined that 10--15 ``core-GHz''
is generally sufficient to reach close to peak performance with a mid- to high-end GPU
in typical biomolecular simulation workloads like the ones used here.
If there is however significant work left for the CPU after offloading the non-bonded and PME computation
(e.g.\ a large amount of bonded interactions or a free energy computation setup),
more CPU cores may be required for an optimal balance.
Additionally, this balance does of course shift as bigger and faster GPUs become available, like the
Tesla V100 (or the similarly performing GeForce RTX 2080Ti) does in fact need around 8--10 cores
equivalent to 16--20 ``core-GHz'' before the performance curve flattens (purple lines).
The increasing size of GPUs however also poses a computational challenge:
large devices are difficult to saturate and can not obtain their
peak throughput with common workloads like the MEM benchmark which is why
the advantage of the Tesla V100 over the RTX 2080
is relatively small especially in comparison to the much larger RIB benchmark case
(purple and dark green lines on Fig.~\ref{fig:perfVsCores}).

An additional benefit of PME offload is that the achievable peak performance also increases and, with fast GPUs,
a significant performance increase is achieved (see dark green and purple lines in Fig.~\ref{fig:perfVsCores})
that was previously not possible with slower accompanying CPUs.
Conversely, however, with CPU-GPU setups ideal for earlier \gromacs versions, PME offload may not improve performance.
In particular on legacy GPU hardware, PME offload is often slower what is reflected in the early performance cross-over
around 5--6 cores per GPU in the Xeon E5-2620v3 CPUs with Tesla K80 benchmarks (solid and dotted pink lines).

In summary, the 2018 version, thanks to the combination of the dynamic pruning algorithm and PME offload, enables using
fewer and/or slower CPU cores in combination with faster GPUs, which is particularly suitable for throughput studies on
GPU-dense hardware.
With hardware setups close to balanced for the 2016 release,
PME offload will not lead to significant performance improvements but the new code capabilities
open up the possibility for upgrades with further accelerators.
As an example, given an 8-core workstation CPU (like an AMD Ryzen 7 2700), which will be approximately
as fast as 10--16 cores in Fig.~\ref{fig:perfVsCores}, when combined with a GTX 1080, there would be 
little improvement from PME offload, and even with an RTX 2080, the improvement would be modest
(assuming similar workload as ours).
However, in such a workstation, adding a second GPU would nearly double the performance.

\subsection{Which hardware is optimal for MD?}

\begin{table}[htbp]
\caption{Single-node performances (average over two runs) for \gromacs 2018 and corresponding P/P ratios.
On nodes with $N$ GPUs, the aggregate performance of $N$ simulations is reported (except 4R where 4 simulations are run).\\
\footnotesize
U = rack space requirements in units per node, D for desktop chassis.
\cuEight using  CUDA 8.0 + GCC 5.4 + Intel MPI 2017,
\cuNine using CUDA 9.1 + GCC 6.4 + Intel MPI 2017,
\mpcdf using CUDA 10.0 + GCC 6.4 + Intel MPI 2018.
P/P ratios were normalized such that values $\geq 1$ result.
}
\label{tab:numbers2018}
\begin{center}
\vspace{-10mm}
\footnotesize
\begin{spreadtab}{{tabular}{clcclSSN{5}{0}SS>{\bfseries}l}}
@   & @                        & @            & @      & @                & @                 & @               & @                                                                                          & @\mult{P/P MEM}       & @\mult{P/P RIB}         & @\mult{P/P}             \\ 
@U  & @processor               & @sockets \mal& @clock & @mounted         & @\mult{MEM}       & @\mult{RIB}     & @\mult{{$\approx$price}}                                                                   & @\mult{(\nsday/}      & @\mult{(\nsday/}        & @\mult{}                \\ 
@   & @AMD/Intel               & @ cores      & @(GHz) & @GPUs            & @\mult{(\nsday)}  & @\mult{(\nsday)}& @\mult{{(\euro\ net)}}                                                                     & @\mult{\fMEM\,\euro)} & @\mult{\fRIB\,\euro)}   & @\mult{}                \\ \midrule
@ 1 & @ E3-1270v5     \cuEight & @ 1 \mal  4  & @3.6   & @1070            &     61.429 *\fOld &   3.061 *\fOld  & \eurRam+\eurSsd+\eurBoardEThree+\eurChassis+\eurCpuFourCoreSkylake+\eurGtxTenSeventy       & [-3,0]/([-1,0]/\fMEM) & [-3,0]/([-2,0]/\fRIB)   & \SThidecol              \\
@ 1 & @ E3-1270v5     \cuEight & @ 1 \mal  4  & @3.6   & @1070Ti          &     68.3385*\fOld &   3.148 *\fOld  & \eurRam+\eurSsd+\eurBoardEThree+\eurChassis+\eurCpuFourCoreSkylake+\eurGtxTenSeventyTi     & [-3,0]/([-1,0]/\fMEM) & [-3,0]/([-2,0]/\fRIB)   & 2/3*[-1,0] + 1/3*[-2,0] \\
@ 1 & @ E3-1270v5     \cuEight & @ 1 \mal  4  & @3.6   & @1080            &     69.425 *\fOld &   3.0245*\fOld  & \eurRam+\eurSsd+\eurBoardEThree+\eurChassis+\eurCpuFourCoreSkylake+\eurGtxTenEighty        & [-3,0]/([-1,0]/\fMEM) & [-3,0]/([-2,0]/\fRIB)   & 2/3*[-1,0] + 1/3*[-2,0] \\
@ 1 & @ E3-1270v5     \cuEight & @ 1 \mal  4  & @3.6   & @1080Ti          &     82.5705*\fOld &   3.704 *\fOld  & \eurRam+\eurSsd+\eurBoardEThree+\eurChassis+\eurCpuFourCoreSkylake+\eurGtxTenEightyTi      & [-3,0]/([-1,0]/\fMEM) & [-3,0]/([-2,0]/\fRIB)   & 2/3*[-1,0] + 1/3*[-2,0] \\ \midrule
@ 1 & @ E3-1240v6     \cuNine  & @ 1 \mal  4  & @3.7   & @ --             &     11.6965       &   0.6915        & \eurRam+\eurSsd+\eurBoardCoreiThree+\eurChassis+\eurCpuFourCoreKaby                        & [-3,0]/([-1,0]/\fMEM) & [-3,0]/([-2,0]/\fRIB)   & 2/3*[-1,0] + 1/3*[-2,0] \\
@ 1 & @ E3-1240v6     \cuEight & @ 1 \mal  4  & @3.7   & @1070            &     67.0475*\fOld &   3.392 *\fOld  & \eurRam+\eurSsd+\eurBoardCoreiThree+\eurChassis+\eurCpuFourCoreKaby+\eurGtxTenSeventy      & [-3,0]/([-1,0]/\fMEM) & [-3,0]/([-2,0]/\fRIB)   & 2/3*[-1,0] + 1/3*[-2,0] \\
@ 1 & @ E3-1240v6     \cuEight & @ 1 \mal  4  & @3.7   & @1080            &     76.138 *\fOld &   3.395 *\fOld  & \eurRam+\eurSsd+\eurBoardCoreiThree+\eurChassis+\eurCpuFourCoreKaby+\eurGtxTenEighty       & [-3,0]/([-1,0]/\fMEM) & [-3,0]/([-2,0]/\fRIB)   & 2/3*[-1,0] + 1/3*[-2,0] \\
@ 1 & @ E3-1240v6     \cuEight & @ 1 \mal  4  & @3.7   & @1080Ti          &     93.1235*\fOld &   4.2695*\fOld  & \eurRam+\eurSsd+\eurBoardCoreiThree+\eurChassis+\eurCpuFourCoreKaby+\eurGtxTenEightyTi     & [-3,0]/([-1,0]/\fMEM) & [-3,0]/([-2,0]/\fRIB)   & 2/3*[-1,0] + 1/3*[-2,0] \\
@ 1 & @ E3-1240v6     \cuNine  & @ 1 \mal  4  & @3.7   & @2080            &    110.6905       &   4.85          & \eurRam+\eurSsd+\eurBoardCoreiThree+\eurChassis+\eurCpuFourCoreKaby+\eurRtxTwEighty        & [-3,0]/([-1,0]/\fMEM) & [-3,0]/([-2,0]/\fRIB)   & 2/3*[-1,0] + 1/3*[-2,0] \\ 
@ 1 & @ E3-1240v6     \cuNine  & @ 1 \mal  4  & @3.7   & @2080Ti          &    133.259        &   5.766         & \eurRam+\eurSsd+\eurBoardCoreiThree+\eurChassis+\eurCpuFourCoreKaby+\eurRtxTwEightyTi      & [-3,0]/([-1,0]/\fMEM) & [-3,0]/([-2,0]/\fRIB)   & 2/3*[-1,0] + 1/3*[-2,0] \\ \midrule 
@ 1 & @ Core i7-6700K \cuEight & @ 1 \mal  4  & @4.0   & @1070            &     66.0415*\fOld &   3.2025*\fOld  & \eurRam+\eurSsd+\eurBoardCoreiThree+\eurChassis+\eurCpuFourCoreiSeven+\eurGtxTenSeventy    & [-3,0]/([-1,0]/\fMEM) & [-3,0]/([-2,0]/\fRIB)   & 2/3*[-1,0] + 1/3*[-2,0] \\
@ 1 & @ Core i7-6700K \cuEight & @ 1 \mal  4  & @4.0   & @1080Ti          &     90.2590*\fOld &   3.9305*\fOld  & \eurRam+\eurSsd+\eurBoardCoreiThree+\eurChassis+\eurCpuFourCoreiSeven+\eurGtxTenEightyTi   & [-3,0]/([-1,0]/\fMEM) & [-3,0]/([-2,0]/\fRIB)   & 2/3*[-1,0] + 1/3*[-2,0] \\ \midrule
@ 1 & @ Silver 4110   \cuEight & @ 1 \mal  8  & @2.1   & @1080Ti          &     95.2495*\fOld &   5.1415*\fOld  & \eurRam+\eurSsd+\eurBoardSkylake+\eurChassis+\eurCpuEightCoreSkylake+\eurGtxTenEightyTi    & [-3,0]/([-1,0]/\fMEM) & [-3,0]/([-2,0]/\fRIB)   & 2/3*[-1,0] + 1/3*[-2,0] \\
@ 1 & @ Silver 4110   \cuEight & @ 1 \mal  8  & @2.1   & @1080\two        &    127.97  *\fOld &   6.805 *\fOld  & \eurRam+\eurSsd+\eurBbSkylake+\eurCpuEightCoreSkylake+2*\eurGtxTenEighty                   & [-3,0]/([-1,0]/\fMEM) & [-3,0]/([-2,0]/\fRIB)   & 2/3*[-1,0] + 1/3*[-2,0] \\
@ 1 & @ Silver 4110   \cuEight & @ 1 \mal  8  & @2.1   & @1080Ti\two      &    151.866 *\fOld &   8.627 *\fOld  & \eurRam+\eurSsd+\eurBbSkylake+\eurCpuEightCoreSkylake+2*\eurGtxTenEightyTi                 & [-3,0]/([-1,0]/\fMEM) & [-3,0]/([-2,0]/\fRIB)   & 2/3*[-1,0] + 1/3*[-2,0] \\ \midrule
@ 1 & @ E5-2630v4     \cuEight & @ 1 \mal 10  & @2.2   & @1070            &     71.813 *\fOld &   3.743 *\fOld  & \eurRam+\eurSsd+\eurBoardBroadwell+\eurChassis+\eurCpuTenCoreBroadwell+\eurGtxTenSeventy   & [-3,0]/([-1,0]/\fMEM) & [-3,0]/([-2,0]/\fRIB)   & 2/3*[-1,0] + 1/3*[-2,0] \\
@ 1 & @ E5-2630v4     \cuEight & @ 1 \mal 10  & @2.2   & @1070Ti          &     80.3995*\fOld &   3.9215*\fOld  & \eurRam+\eurSsd+\eurBoardBroadwell+\eurChassis+\eurCpuTenCoreBroadwell+\eurGtxTenSeventyTi & [-3,0]/([-1,0]/\fMEM) & [-3,0]/([-2,0]/\fRIB)   & 2/3*[-1,0] + 1/3*[-2,0] \\
@ 1 & @ E5-2630v4     \cuEight & @ 1 \mal 10  & @2.2   & @1080            &     81.3885*\fOld &   3.733 *\fOld  & \eurRam+\eurSsd+\eurBoardBroadwell+\eurChassis+\eurCpuTenCoreBroadwell+\eurGtxTenEighty    & [-3,0]/([-1,0]/\fMEM) & [-3,0]/([-2,0]/\fRIB)   & 2/3*[-1,0] + 1/3*[-2,0] \\
@ 1 & @ E5-2630v4     \cuEight & @ 1 \mal 10  & @2.2   & @1080Ti          &    101.472 *\fOld &   4.869 *\fOld  & \eurRam+\eurSsd+\eurBoardBroadwell+\eurChassis+\eurCpuTenCoreBroadwell+\eurGtxTenEightyTi  & [-3,0]/([-1,0]/\fMEM) & [-3,0]/([-2,0]/\fRIB)   & 2/3*[-1,0] + 1/3*[-2,0] \\
@ 1 & @ E5-2630v4     \cuNine  & @ 1 \mal 10  & @2.2   & @2080            &    115.409        &   5.8795        & \eurRam+\eurSsd+\eurBoardBroadwell+\eurChassis+    \eurCpuTenCoreBroadwell+\eurRtxTwEighty & [-3,0]/([-1,0]/\fMEM) & [-3,0]/([-2,0]/\fRIB)   & 2/3*[-1,0] + 1/3*[-2,0] \\
@ 1 & @ E5-2630v4     \cuNine  & @ 1 \mal 10  & @2.2   & @2080Ti          &    146.2805       &   7.271         & \eurRam+\eurSsd+\eurBoardBroadwell+\eurChassis+  \eurCpuTenCoreBroadwell+\eurRtxTwEightyTi & [-3,0]/([-1,0]/\fMEM) & [-3,0]/([-2,0]/\fRIB)   & 2/3*[-1,0] + 1/3*[-2,0] \\
@ 1 & @ E5-2630v4     \cuEight & @ 1 \mal 10  & @2.2   & @1080Ti\two      &    175.1305*\fOld &   8.375 *\fOld  & \eurRam+\eurSsd+\eurBoardBroadwell+\eurChassis+320+\eurCpuTenCoreBroadwell+2*\eurGtxTenEightyTi& [-3,0]/([-1,0]/\fMEM) & [-3,0]/([-2,0]/\fRIB)   & 2/3*[-1,0] + 1/3*[-2,0] \\
@ 1 & @ E5-2630v4     \cuNine  & @ 1 \mal 10  & @2.2   & @2080\two        &    201.331        &  10.126         & \eurRam+\eurSsd+\eurBoardBroadwell+\eurChassis+320+\eurCpuTenCoreBroadwell+2*\eurRtxTwEighty& [-3,0]/([-1,0]/\fMEM) & [-3,0]/([-2,0]/\fRIB)   & 2/3*[-1,0] + 1/3*[-2,0] \\ \midrule
@ 1 & @ Silver 4114   \cuEight & @ 1 \mal 10  & @2.2   & @1070Ti          &     79.323 *\fOld &   4.3815*\fOld  & \eurRam+\eurSsd+\eurBoardSkylake+\eurChassis+\eurCpuTenCoreSkylake+\eurGtxTenSeventyTi     & [-3,0]/([-1,0]/\fMEM) & [-3,0]/([-2,0]/\fRIB)   & 2/3*[-1,0] + 1/3*[-2,0] \\
@ 1 & @ Silver 4114   \cuEight & @ 1 \mal 10  & @2.2   & @1080            &     80.8655*\fOld &   4.235 *\fOld  & \eurRam+\eurSsd+\eurBoardSkylake+\eurChassis+\eurCpuTenCoreSkylake+\eurGtxTenEighty        & [-3,0]/([-1,0]/\fMEM) & [-3,0]/([-2,0]/\fRIB)   & 2/3*[-1,0] + 1/3*[-2,0] \\
@ 1 & @ Silver 4114   \cuEight & @ 1 \mal 10  & @2.2   & @1080Ti          &    101.3685*\fOld &   5.439 *\fOld  & \eurRam+\eurSsd+\eurBoardSkylake+\eurChassis+\eurCpuTenCoreSkylake+\eurGtxTenEightyTi      & [-3,0]/([-1,0]/\fMEM) & [-3,0]/([-2,0]/\fRIB)   & 2/3*[-1,0] + 1/3*[-2,0] \\ 
@ 1 & @ Silver 4114   \cuNine  & @ 1 \mal 10  & @2.2   & @2080Ti          &    147.292        &   7.851         & \eurRam+\eurSsd+\eurBoardSkylake+\eurChassis+\eurCpuTenCoreSkylake+\eurRtxTwEightyTi       & [-3,0]/([-1,0]/\fMEM) & [-3,0]/([-2,0]/\fRIB)   & 2/3*[-1,0] + 1/3*[-2,0] \\ 
@ 1 & @ Silver 4114   \cuEight & @ 1 \mal 10  & @2.2   & @1080\two        &    139.2165*\fOld &   7.1135*\fOld  & \eurRam+\eurSsd+\eurBbSkylake+\eurCpuTenCoreSkylake+2*\eurGtxTenEighty                     & [-3,0]/([-1,0]/\fMEM) & [-3,0]/([-2,0]/\fRIB)   & 2/3*[-1,0] + 1/3*[-2,0] \\
@ 1 & @ Silver 4114   \cuEight & @ 1 \mal 10  & @2.2   & @1080Ti\two      &    165.064 *\fOld &   9.3465*\fOld  & \eurRam+\eurSsd+\eurBbSkylake+\eurCpuTenCoreSkylake+2*\eurGtxTenEightyTi                   & [-3,0]/([-1,0]/\fMEM) & [-3,0]/([-2,0]/\fRIB)   & 2/3*[-1,0] + 1/3*[-2,0] \\ \midrule 
@ D & @ Ryzen 1950X   \cuNine  & @ 1 \mal 16  & @3.4   & @1080Ti          &     94.8985       &   5.006         & 1830+1*\eurGtxTenEightyTi                                                                  & [-3,0]/([-1,0]/\fMEM) & [-3,0]/([-2,0]/\fRIB)   & 2/3*[-1,0] + 1/3*[-2,0] \\
@ D & @ Ryzen 1950X   \cuNine  & @ 1 \mal 16  & @3.4   & @2080            &    106.08         &   5.556         & 1830+1*\eurRtxTwEighty                                                                     & [-3,0]/([-1,0]/\fMEM) & [-3,0]/([-2,0]/\fRIB)   & 2/3*[-1,0] + 1/3*[-2,0] \\
@ D & @ Ryzen 1950X   \cuNine  & @ 1 \mal 16  & @3.4   & @1080Ti\two      &    172.6205       &   9.162         & 1830+2*\eurGtxTenEightyTi                                                                  & [-3,0]/([-1,0]/\fMEM) & [-3,0]/([-2,0]/\fRIB)   & 2/3*[-1,0] + 1/3*[-2,0] \\
@ D & @ Ryzen 1950X   \cuNine  & @ 1 \mal 16  & @3.4   & @2080\two        &    196.731        &  10.0825        & 1830+2*\eurRtxTwEighty                                                                     & [-3,0]/([-1,0]/\fMEM) & [-3,0]/([-2,0]/\fRIB)   & 2/3*[-1,0] + 1/3*[-2,0] \\
@ D & @ Ryzen 1950X   \cuNine  & @ 1 \mal 16  & @3.4   & @2080\three      &    267.51         &  12.788         & 1830+3*\eurRtxTwEighty                                                                     & [-3,0]/([-1,0]/\fMEM) & [-3,0]/([-2,0]/\fRIB)   & 2/3*[-1,0] + 1/3*[-2,0] \\
@ D & @ Ryzen 1950X   \cuNine  & @ 1 \mal 16  & @3.4   & @2080\four       &    332.825        &  14.066         & 1830+4*\eurRtxTwEighty                                                                     & [-3,0]/([-1,0]/\fMEM) & [-3,0]/([-2,0]/\fRIB)   & 2/3*[-1,0] + 1/3*[-2,0] \\ \midrule
@ 1 & @ Epyc 7401P    \cuNine  & @ 1 \mal 24  & @2.0   & @  --            &     28.706        &   2.2775        & 4250-2*\eurGtxTenSeventyTi                                                                 & [-3,0]/([-1,0]/\fMEM) & [-3,0]/([-2,0]/\fRIB)   & 2/3*[-1,0] + 1/3*[-2,0] \\
@ 1 & @ Epyc 7401P    \cuNine  & @ 1 \mal 24 & @2.0   & @1080Ti\two      &    191.6575        &   9.4905        & 4250+2*\eurGtxTenEightyTi-2*\eurGtxTenSeventyTi                                            & [-3,0]/([-1,0]/\fMEM) & [-3,0]/([-2,0]/\fRIB)   & 2/3*[-1,0] + 1/3*[-2,0] \\
@ 1 & @ Epyc 7401P    \cuNine  & @ 1 \mal 24 & @2.0   & @1080Ti\four     &    369.057         &  16.9585        & 4250+4*\eurGtxTenEightyTi-2*\eurGtxTenSeventyTi                                            & [-3,0]/([-1,0]/\fMEM) & [-3,0]/([-2,0]/\fRIB)   & 2/3*[-1,0] + 1/3*[-2,0] \\ \midrule
@ 2 & @Gold6148F \mal2 \mpcdf  & @ 2 \mal 20 & @2.4   & @V100\two        &    300.755         &  19.9495        & \eurRam+\eurSsd+\eurBbSkylake+2*\eurTeslaVIOO+2*\eurCpuTwCoreSkylake                       & [-3,0]/([-1,0]/\fMEM) & [-3,0]/([-2,0]/\fRIB)   & 2/3*[-1,0] + 1/3*[-2,0] \\
@ 2 & @Gold6148F \mal2 \mpcdf  & @ 2 \mal 20 & @2.4   & @V100\two (4R)   &    393.3195        &  27.2655        & \eurRam+\eurSsd+\eurBbSkylake+2*\eurTeslaVIOO+2*\eurCpuTwCoreSkylake                       & [-3,0]/([-1,0]/\fMEM) & [-3,0]/([-2,0]/\fRIB)   & 2/3*[-1,0] + 1/3*[-2,0] \\
\end{spreadtab}
\end{center}
\end{table}

\newcommand{\rarr}{\rightarrow}
\newcommand{\pCpu}{\mbox{$^\star$}}   

\begin{table}[htbp]
\caption{Single-node performances for \gromacs 2018 as in Table~\ref{tab:numbers2018},
but for upgrading existing nodes with modern GPUs.
Here, the P/P ratios have been calculated from the performance increase
(i.e.\ performance with upgraded GPU(s) minus performance with old GPU)
and the cost of the GPU(s).
All benchmarks use CUDA 9.1, GCC 6.4, and, for the multi-GPU setups, Intel MPI 2017.
For optimum performance, PME mesh part was offloaded to the GPU, except where indicated (\pCpu).
}
\label{tab:upgrade}
\STautoround{1}
\begin{center}
\vspace{-3mm}
\footnotesize
\FPset{\baseMEM}{104.8295}   
\FPset{\baseRIB}{6.7425}     
\FPset{\baseMEMb}{83.2155}  
\FPset{\baseRIBb}{4.9575}   
\FPset{\baseMEMc}{122.921}  
\FPset{\baseRIBc}{6.905}    
\FPset{\baseMEMd}{26.9045}  
\FPset{\baseRIBd}{1.624}    
\begin{spreadtab}{{tabular}{clcclSSN{5}{0}SS>{\bfseries}l}}
@   & @                      & @            & @      & @                & @               & @               & @                       & @\mult{P/P MEM}                   & @\mult{P/P RIB}                   & @\mult{P/P}             \\ 
@U  & @processor             & @sockets \mal& @clock & @mounted         & @\mult{MEM}     & @\mult{RIB}     & @\mult{{$\approx$price}}& @\mult{(\nsday/}                  & @\mult{(\nsday/}                  & @\mult{}                \\ 
@   & @Intel                 & @ cores      & @(GHz) & @GPUs            & @\mult{(\nsday)}& @\mult{(\nsday)}& @\mult{{(\euro\ net)}}  & @\mult{\fMEM\,\euro)}             & @\mult{\fRIB\,\euro)}             & @\mult{}                \\ \midrule
@   & @\multicolumn{8}{c}{existing node with old GPU:}\\
@ 1 & @ E3-1270v2            & @ 1 \mal 4   & @3.5   & @680             &:={26.9045}\pCpu &       1.624     & 0                       & @-                                & @-                                &                         \\
@   & @\multicolumn{8}{c}{with upgraded GPU:}\\
@ 1 & @ E3-1270v2            & @ 1 \mal 4   & @3.5   & @2080            &     91.7075     &       4.013     & \eurRtxTwEighty         & ([-3,0]-\baseMEMd)/([-1,0]/\fMEM) & ([-3,0]-\baseRIBd)/([-2,0]/\fRIB) &                         \\ \midrule
@   & @\multicolumn{8}{c}{existing node with old GPUs:}\\
@ 2 & @ E5-2670v2 \mal2      & @ 2 \mal 10  & @2.5   & @780Ti\two       &:={104.8295}\pCpu&       6.7425    & 0                       & @-                                & @-                                &                         \\
@   & @\multicolumn{8}{c}{with upgraded GPUs:}\\
@ 2 & @ E5-2670v2 \mal2      & @ 2 \mal 10  & @2.5   & @1080\two        &    163.393      &       7.825     & 2*\eurGtxTenEighty      & ([-3,0]-\baseMEM )/([-1,0]/\fMEM) & ([-3,0]-\baseRIB )/([-2,0]/\fRIB) & 2/3*[-1,0] + 1/3*[-2,0] \\
@ 2 & @ E5-2670v2 \mal2      & @ 2 \mal 10  & @2.5   & @1080Ti\two      &    208.4115     &      10.1615    & 2*\eurGtxTenEightyTi    & ([-3,0]-\baseMEM )/([-1,0]/\fMEM) & ([-3,0]-\baseRIB )/([-2,0]/\fRIB) & 2/3*[-1,0] + 1/3*[-2,0] \\
@ 2 & @ E5-2670v2 \mal2      & @ 2 \mal 10  & @2.5   & @1080Ti\four     &    361.2145     &      17.896     & 4*\eurGtxTenEightyTi    & ([-3,0]-\baseMEM )/([-1,0]/\fMEM) & ([-3,0]-\baseRIB )/([-2,0]/\fRIB) & 2/3*[-1,0] + 1/3*[-2,0] \\ \midrule
@   & @\multicolumn{8}{c}{existing node with old GPUs:}\\
@ 2 & @ E5-2680v2 \mal2      & @ 2 \mal 10  & @2.8   & @K20Xm\two       &:={83.2155}\pCpu &       4.9575    & 0                       & @-                                & @-                                & \SThidecol              \\
@   & @\multicolumn{8}{c}{with upgraded GPUs:}\\
@ 2 & @ E5-2680v2 \mal2      & @ 2 \mal 10  & @2.8   & @1080Ti\two      &    212.695      &      10.108     & 2*\eurGtxTenEightyTi    & ([-3,0]-\baseMEMb)/([-1,0]/\fMEM) & ([-3,0]-\baseRIBb)/([-2,0]/\fRIB) & 2/3*[-1,0] + 1/3*[-2,0] \\
@ 2 & @ E5-2680v2 \mal2      & @ 2 \mal 10  & @2.8   & @2080\two        &    237.7035     &      11.454     & 2*\eurRtxTwEighty       & ([-3,0]-\baseMEMb)/([-1,0]/\fMEM) & ([-3,0]-\baseRIBb)/([-2,0]/\fRIB) & 2/3*[-1,0] + 1/3*[-2,0] \\
@ 2 & @ E5-2680v2 \mal2      & @ 2 \mal 10  & @2.8   & @2080\four       &    409.5610     &      20.303     & 4*\eurRtxTwEighty       & ([-3,0]-\baseMEMb)/([-1,0]/\fMEM) & ([-3,0]-\baseRIBb)/([-2,0]/\fRIB) & 2/3*[-1,0] + 1/3*[-2,0] \\
\end{spreadtab}
\end{center}
\end{table}

\begin{figure}
\begin{center}
\includegraphics[width=\textwidth]{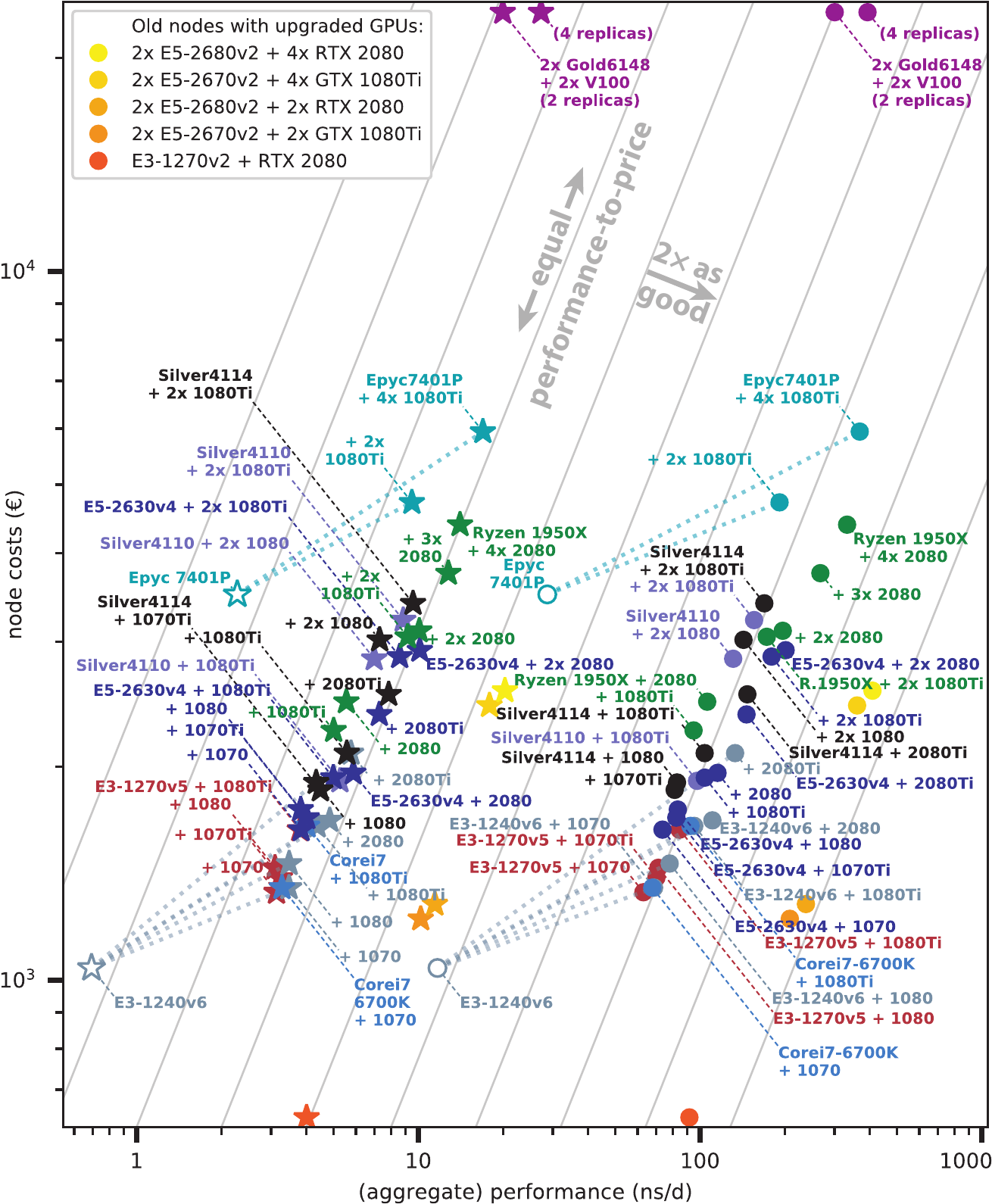}
\caption{(Aggregate) simulation performance in relation to net node costs. 
MEM (circles) and RIB (stars) symbols are colored depending on CPU type.
Symbols with white fill denote nodes without GPU acceleration;
dotted lines connect GPU nodes with their CPU counterparts.
Grey: isolines of equal P/P ratio like in Fig.~\ref{fig:eval2014} with superior configurations to the lower right.
Old nodes with upgraded GPUs from Table~\ref{tab:upgrade} are shown in yellow-orange colors (legend).}
\label{fig:costsVsPerf}
\end{center}
\end{figure}

Table~\ref{tab:numbers2018} and Fig.~\ref{fig:costsVsPerf} show the results of our
current hardware ranking.
Overall, the P/P ratio of the examined consumer GPU nodes 
is about a factor of 3--6 higher compared to their counterparts without GPUs.
The P/P ratios of new nodes with consumer GPUs are all very similar;
most of them are less than a factor of 1.5 apart and thus
scatter about one isoline in the log-log plot (Fig.~\ref{fig:costsVsPerf}).
Note that this similarity results from our hardware preselection and does not imply that
any node with consumer GPU(s) has a comparable P/P ratio.
There are lots of hardware combinations possible that we did not consider
because high costs of one or more individual components preclude a competitive P/P ratio from the start.

The cheapest nodes with a good P/P ratio, starting at $\approx$1,400 \Euro net,
are Intel E3-1270v5, E3-1240v6, or Core i7-6700K CPUs 
combined with a GeForce 1070(Ti), respectively.
The best P/P ratio (with current pricing) is offered by combining a E5-2630v4 or Ryzen 1950X CPU 
with two (or possibly more) RTX 2080 GPUs
starting at 3,000 \Euro net.
The best aggregate performance for consumer GPU nodes was identified for the AMD Epyc 24-core node combined
with four 1080Ti GPUs.

Concerning space requirements,
most node types listed in Table~\ref{tab:numbers2018} fit in one height unit (U) of rack space.
One exception is the Ryzen Threadripper 1950X that was available in a desktop chassis only
(using $\approx$ 4 U, if mounted in a rack).

\subsection{Alternative: upgrade existing nodes with recent GPUs}
An attractive alternative to replacing old GPU nodes with new hardware is 
to solely exchange the existing GPUs with more powerful recent models.
Due to offloading even more interaction types to the GPU,
compared to older versions, \gromacs 2018 demands more compute power
on the GPU side, but less on the CPU side. 
As a result, CPU models from a few years ago often ideally combine with modern GPUs.

For instance, the performance gain for a dual ten-core node with two K20Xm GPUs that was part of the 2014 investigation
is a factor of 3.5 for the MEM benchmark for switching the old GPUs with recent RTX 2080 models.
Table~\ref{tab:upgrade} lists the performance gains for different upgrade scenarios.
The top line in each section shows the performance of an old node with \gromacs 2018,
whereas the following lines show how performance increases when GPUs are upgraded.
Depending on the exact old and new GPU type,
one can easily achieve a twofold higher aggregate performance from these nodes
for the comparatively small investment of just having to buy the GPUs and not whole nodes.
Performance-wise it makes no difference whether four 1080Ti GPUs are combined with
a new EPYC 7401 processor (24 cores), or with two old E5-2670v2 processors (2 \mal 10 cores).
The yellow and orange symbols in Fig.~\ref{fig:costsVsPerf} show the performance
of these nodes in relation to the costs for upgrading them with modern GPUs.

\subsection{Energy efficiency}

\FPset\yrs{5}
\FPset\EpkWh{0.2}

\begin{figure}
\begin{center}
\includegraphics[width=\textwidth]{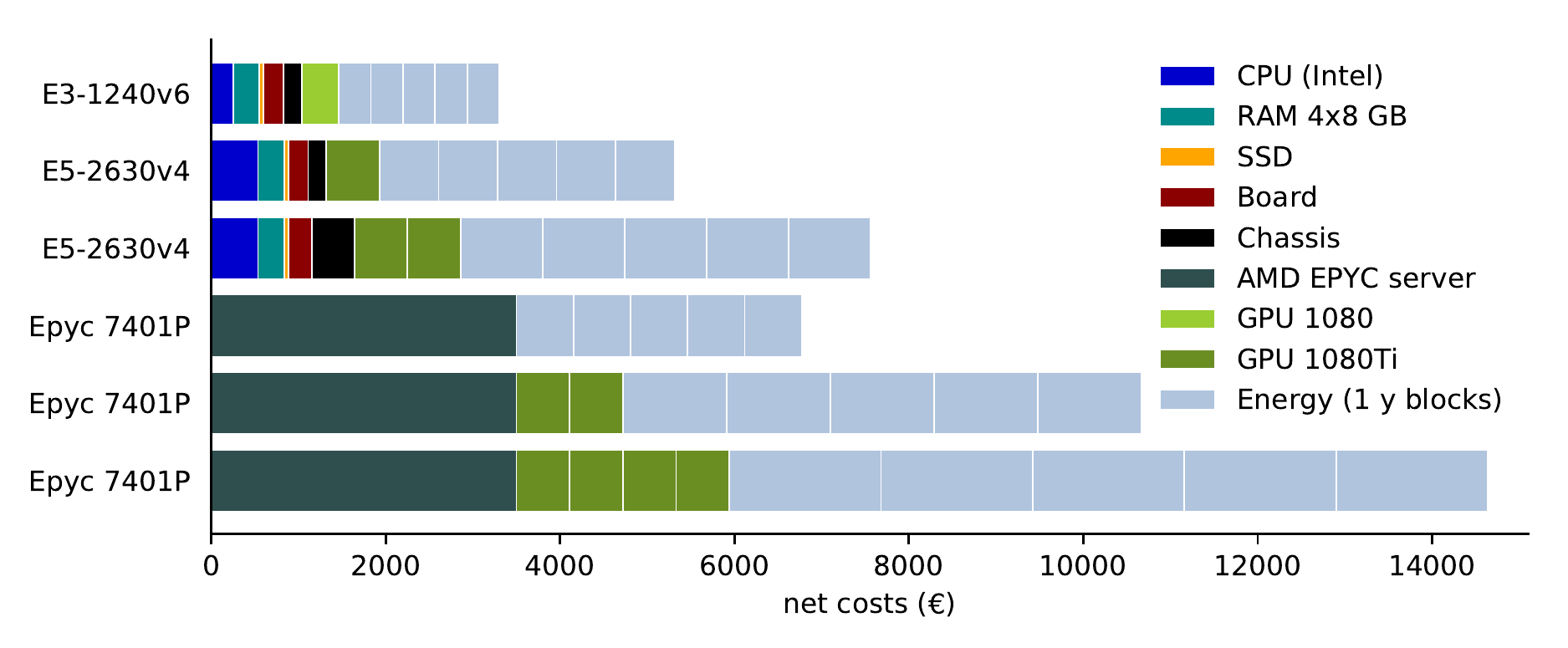}
\caption{Breakdown of total costs for selected node types, 
taking into account 0.2 \Euro per kWh for energy and cooling, 
for a lifetime of five years.}
\label{fig:nodeCosts_energy}
\end{center}
\end{figure}
To establish the total costs of a node over its lifetime,
we determined its power consumption.
Assuming a net cost of 0.2 \Euro per kWh for energy and cooling,
we calculated the total costs of selected node types
as the sum of hardware and energy costs, over five years of operation
(Fig.~\ref{fig:nodeCosts_energy}, five separate one year blocks given for energy and cooling costs).
For the considered node types and an average lifetime of 3--5 years,
the costs for hardware and energy are similar.
This will however not generally be the case,
e.g.\ with professional GPUs, the costs for hardware can easily be three times as high.

\begin{figure}
\begin{center}
\includegraphics[width=\textwidth]{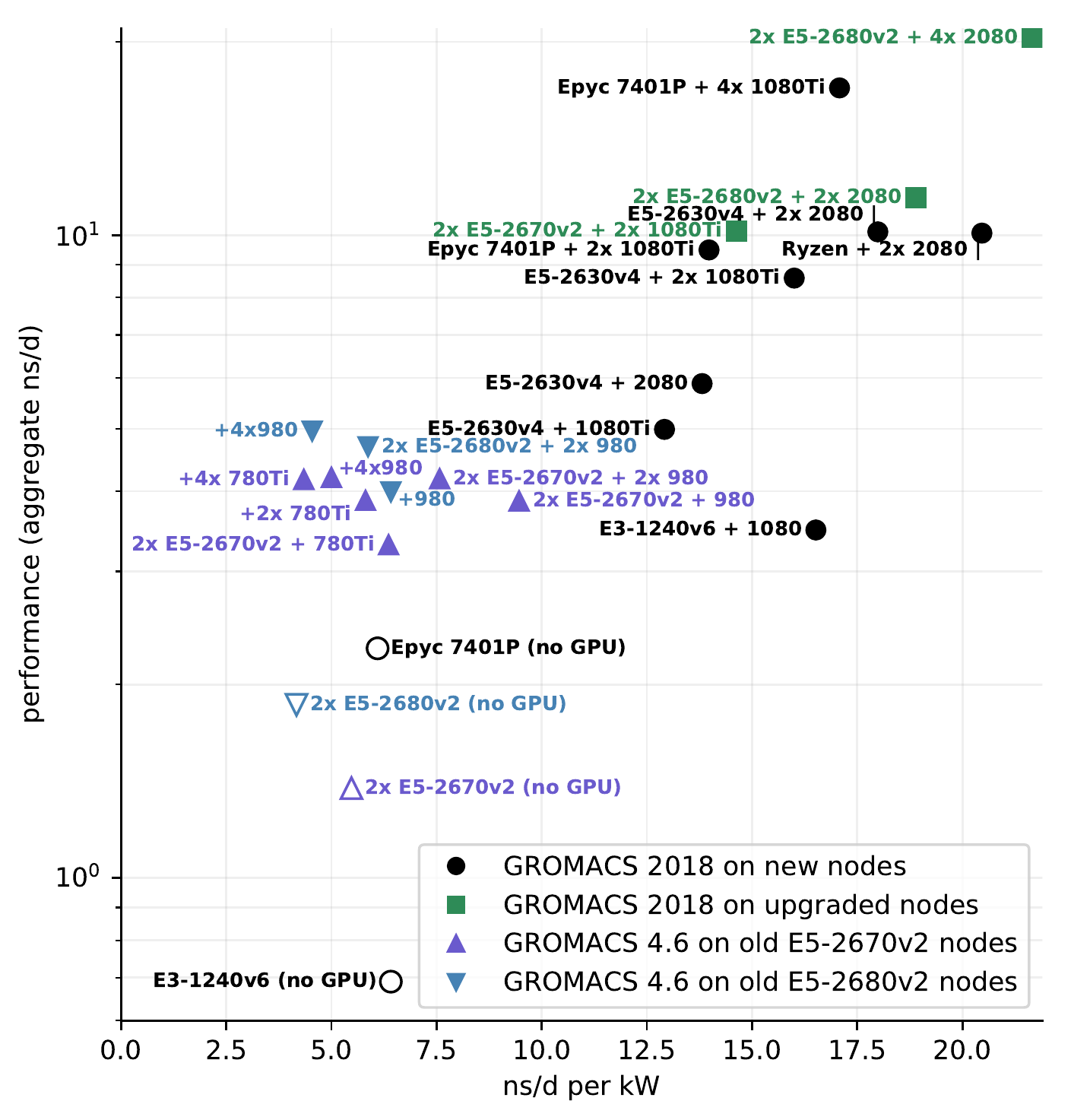}
\caption{RIB performance of selected node types in relation to their energy efficiency.
Nodes without GPUs (white fill) show both low performance as well as low energy efficiency, 
independent of \gromacs version and CPU generation.
Best energy efficiency is recorded for \gromacs 2018 in combination with new GPUs (black and green filled symbols).
}
\label{fig:perfPerWatt}
\end{center}
\end{figure}
Fig.~\ref{fig:perfPerWatt} shows the energy efficiency of selected node types
in relation to their GROMACS performance.
With \gromacs 2018, in addition to their considerably higher simulation performance,
GPU nodes deliver more than two times the performance per Watt 
compared to CPU nodes.

\begin{figure}
\begin{center}
\includegraphics[width=0.9\textwidth]{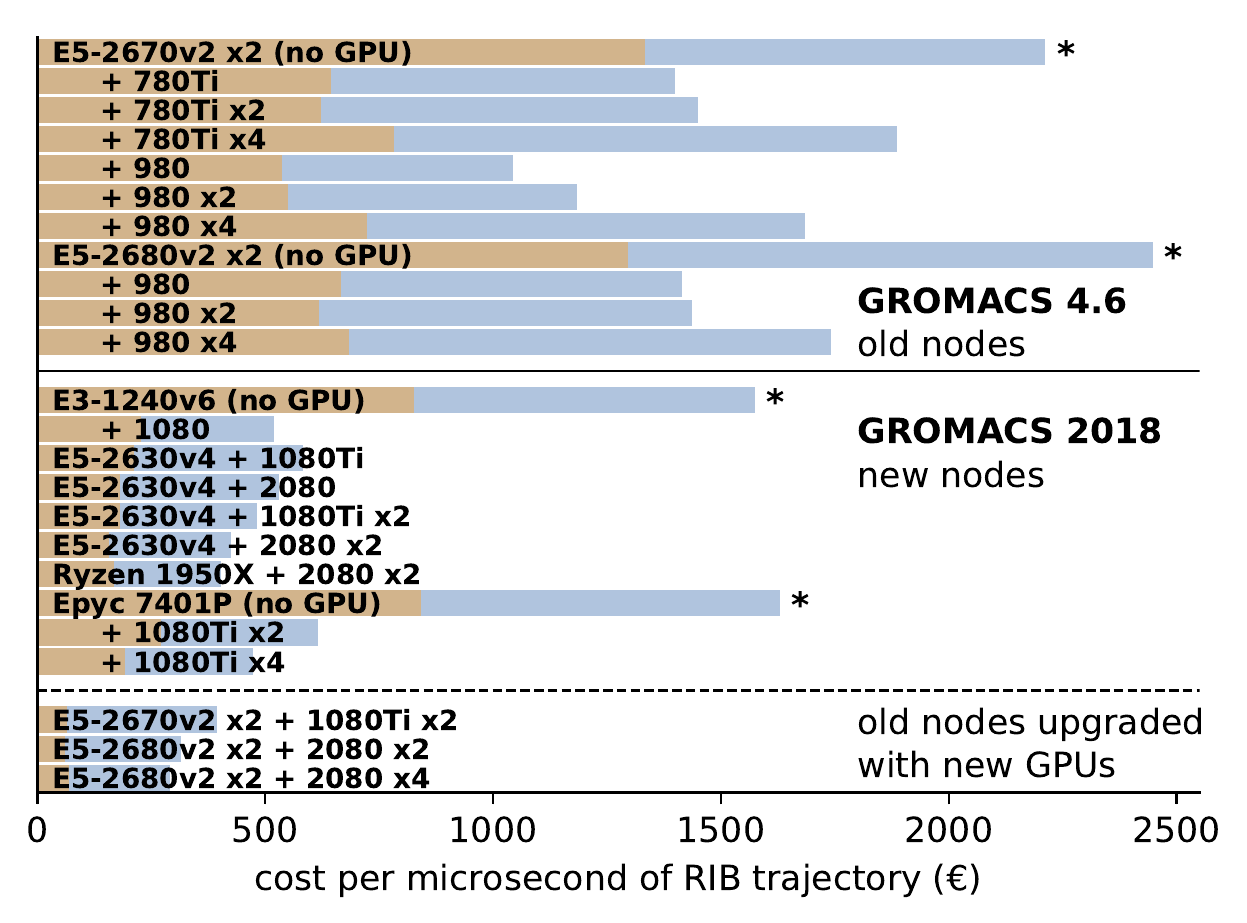}
\caption{RIB trajectory costs for selected node types assuming five years of operation,
including costs of 0.2 \Euro per kWh for energy and cooling.
The top part shows results from 2014 using \gromacs 4.6,\cite{escidoc:2180273}
the lower part depicts results using \gromacs 2018 on recent hardware
and on old hardware that was upgraded with new GPUs (lowermost three bars).
Nodes without GPUs have the highest trajectory production costs (asterisks).}
\label{fig:trajectoryCosts}
\end{center}
\end{figure}
We derive the total costs for producing MD trajectories on different node types
by putting the hardware and energy costs in relation to the amount of produced trajectory
(Fig.~\ref{fig:trajectoryCosts}).
We make three main observations:
(i) Trajectory costs are highest on nodes without consumer GPUs (these are marked with asterisks in the Figure).
(ii) For \gromacs 4.6 on hardware of 2014,
trajectory costs on the best GPU nodes are 0.5--0.6 times that of their CPU-only counterparts.
(iii) With \gromacs 2018 and current hardware,
this factor is reduced to about 0.3. 

\subsection{Coarse grain models}
\begin{figure}
\begin{center}
\includegraphics[width=.9\textwidth]{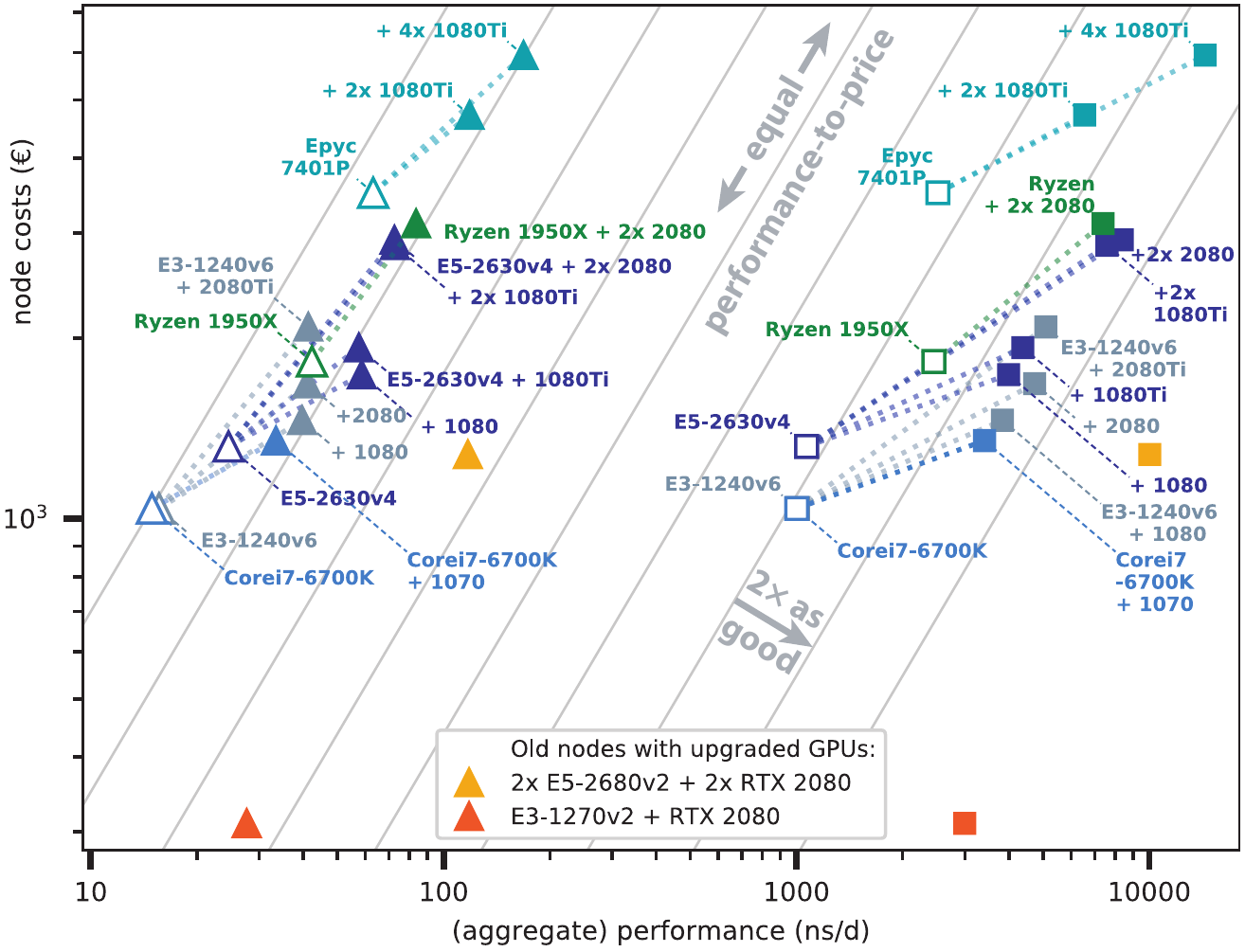}
\caption{(Aggregate) simulation performance in relation to net node costs
as in Fig.~\ref{fig:costsVsPerf}, but for the two coarse grain benchmarks 
(the vesicle and the big membrane patch) using the Martini force field. 
VES (squares) and BIG (triangles) symbols are colored depending on CPU type.
Symbols with white fill denote nodes without GPU acceleration;
dotted lines connect GPU nodes with their CPU counterparts.
Re-used old nodes with new GPUs are shown in orange colors (legend).}
\label{fig:costsVsPerfMartini}
\end{center}
\end{figure}

The Martini systems were run on a subset of the node types used for the
atomistic benchmarks, Fig.~\ref{fig:costsVsPerfMartini} shows the results. 
The overall picture is quite similar to the atomistic benchmarks (Fig.~\ref{fig:costsVsPerf})
but there are some differences as well.

As for the atomistic systems, the P/P ratio is significantly
higher for nodes with consumer GPUs than for CPU nodes. However, the
gap between CPU nodes and consumer GPU nodes is less pronounced than in the
atomistic case. It is about a factor of 2--4 for the small VES system and a factor
of 1.5--2 for the BIG membrane in terms of P/P.
As in the atomistic case, re-used old nodes upgraded with up-to-date consumer GPUs
have the best P/P ratios. 

Although the workload of a Martini coarse grain MD system is quite different
from the workload of an atomistic system (lower particle density, no PME mesh),
it turns out that the node types that are optimal for atomistic MD are
also very well suited for running coarse grain simulations.

\section{Conclusions for \gromacs 2018}
In 2014, we found that nodes with consumer GPUs provide
the \emph{best bang for your buck} due to their significantly higher trajectory output
per invested Euro compared to nodes without GPUs or nodes with professional GPUs.
This applies equally to \gromacs 2018 on current hardware. Moreover,
the existing gap has considerably widened:
Taking into account raw node prices,
today with \gromacs 2018 one can get a factor of three to six times more trajectory on consumer GPU nodes as compared to 
a factor of two to three in 2014 with \gromacs 4.6.
When including costs for energy and cooling, this factor has increased from two to about three.

This marked improvement became possible by offloading also the PME mesh computations to the GPU,
in addition to the short-ranged non-bonded interactions.
PME offloading moves the optimal hardware balance even more towards consumer GPUs.
CPU/GPU combinations identified as optimal in P/P ratio require about four to eight CPU cores per 1080Ti or 2080;
a generally useful rule-of-thumb is that for similar simulation systems as the ones shown here,
10--15 ``core-GHz'' are sufficient and 15--20 ``core-GHz'' are also future-proof for upgrades (or better suited
for workloads with additional CPU computation).

Additionally, PME offloading offers the possibility to cheaply upgrade GPU nodes once tailored for older \gromacs versions.
By keeping everything but exchanging the old GPUs by state-of-the-art models,
an optimal CPU/GPU balance can be restored for \gromacs 2018,
at the comparatively small investment for GPUs only.

\section{Outlook}
Since hardware is continuously evolving and new components 
(CPUs, GPUs, barebones, boards, etc.) will become available in future, 
it is likely that configurations with an even higher P/P ratio than identified in this paper will appear.
Readers who have access to hardware configurations that were not covered here
are encouraged to download our CC-licensed benchmark input files from
\url{https://www.mpibpc.mpg.de/grubmueller/bench} to perform their own benchmarks
such that we can include these data into updated versions of the Tables.

It is worth noting that the presented results do transfer very well to the \gromacs 2019
code, just released at the time of writing this paper. On the performance front, this release
has only modest additions with a few notable exceptions only.
This release introduces PME offload support using OpenCL, which is particularly useful on AMD GPUs,
especially in light of how favorably the (now previous-generation) Radeon Vega GPUs compare
to the competition. Their advantage is particularly pronounced when comparing their P/P ratio against Tesla GPUs
(see Figs.~\ref{fig:kernelPerf} and \ref{fig:gpuPerf2Price}).

The additional feature of the 2019 release worth noting is the ability to offload (most)
bonded interactions with CUDA. However, as \gromacs has highly optimized SIMD kernels for
bonded interactions, this feature will have a beneficial performance impact only in cases where 
either available CPU resources are low or the simulation system contains a significant amount
of bonded work.
For our benchmarks, that would mean that the cases where just a few cores are paired with fast
GPUs would be improved, as the 1--3 core range of the purple and dark green lines in Fig.~\ref{fig:perfVsCores} indicate.

\subsection*{Acknowledgments}
We thank Petra Kellers for thoroughly reading the manuscript; her suggestions
led to numerous improvements.
We thank the MPCDF, especially Markus Rampp and Hermann Lederer, for general help and
for providing some of the hardware that has been benchmarked.
This study was supported by the DFG priority programme
\emph{Software for Exascale Computing} (SPP 1648) and
by the BioExcel CoE (\url{www.bioexcel.eu}), 
a project funded by the European Union contract H2020-EINFRA-2015-1-675728,
the SSF Infrastructure Fellow programme, and the Swedish e-Science Research Centre (SeRC).


\providecommand{\latin}[1]{#1}
\makeatletter
\providecommand{\doi}
  {\begingroup\let\do\@makeother\dospecials
  \catcode`\{=1 \catcode`\}=2 \doi@aux}
\providecommand{\doi@aux}[1]{\endgroup\texttt{#1}}
\makeatother
\providecommand*\mcitethebibliography{\thebibliography}
\csname @ifundefined\endcsname{endmcitethebibliography}
  {\let\endmcitethebibliography\endthebibliography}{}

\end{document}